%% file: RCRspecA.tex
\documentclass[
 aps,
 11pt,
 final,
 notitlepage,
 oneside,
 onecolumn
 nobibnotes,
 nofootinbib,
 superscriptaddress,
 noshowpacs,
 centertags]
{revtex4}

\input{sao_cmd_author.tex}

\begin{document}
\selectlanguage{english}

\title{Study of RCR Catalogue Radio Source Integral Spectra}

\author{\firstname{O.P.}~\surname{Zhelenkova}}
\email{zhe@sao.ru}
\affiliation{\saoname}
\affiliation{ITMO University, Saint Petersburg, 197101, Russia}

\author{\firstname{E.K.}~\surname{Majorova}}
\affiliation{\saoname}

\received{February 21, 2018} \revised{May 15, 2018}

\begin{abstract}
We present the characteristics of the sources found on the averaged scans
of the ``Cold'' experiment 1980--1999 surveys in the right-ascension interval
$2^h$< RA <$7^h$. Thereby, a refinement of the parameters of the RC catalog
sources (RATAN Cold) for this interval is complete. To date, the RCR catalog
(RATAN Cold Refined) covers the right-ascension interval $2^h$< RA <$17^h$ and
includes 830 sources. The spectra are built for them with the use of new
data in the range of 70--230 MHz. The dependence between the spectral indices
$\alpha_{0.5}$, $\alpha_{3.94}$ and integral flux density at the frequencies
of 74 and 150 MHz, at 1.4, 3.94 and 4.85 GHz is discussed. We found that at
150 MHz in most sources the spectral index $\alpha_{0.5}$ gets steeper with
increasing flux density.
In general, the sources with flat spectra are weaker in terms of flux density
than the sources with steep spectra, which especially differs at 150 MHz.
We believe that this is due to the brightness of their extended components,
which can be determined by the type of accretion and the neighborhood of
the source.
\end{abstract}

\maketitle

\section{Introduction}
Within the ``Cold'' experiment~\cite{h,ber,pa1} from 1980 to 2000,
several surveys of the sky strip on the declination of the microquasar
SS\,433 (\mbox{$Dec_{2000}=4^{\circ}59'\pm20'$}) performed at the
RATAN-600 radio telescope.
Based on the 1980--1981 observations, the RC catalog of radio sources
(RATAN\,Cold)~\cite{pa2,pa3} was published, including 1209 objects with
the flux density higher than 5\,mJy, detected at the frequency of 3.94\,GHz.
At such flux densities, practically all the sources found within the ``Cold''
experiment are galaxies with active nuclei (AGN, Active Galaxy
Nucleus) and powerful sources in the radio range~\cite{zhe}.

In AGN, synchrotron radiation in the radio range is produced by
ultrarelativistic particles in the magnetic field. High-energy cosmic
particles are accelerated in the jet that transports them from the active
nucleus to the shock region, called the hot spot~\cite{meis}.
Then a beam of cosmic rays expands from the center of the shock region,
forming extended components of radio emission~\cite{krause}.

Synchrotron radiation in all the processes occurring in different structures
of a radio source is usually described by a power function
$F(\nu)\propto\nu^{\alpha}$, where $F(\nu)$ is the flux density at
the $\nu$ frequency. However, internal variations in the energy distribution
of cosmic rays, the environment of the source, the processes taking place in
the interstellar medium, spectral ``aging''~\cite{harwood}, caused by a
faster de-excitation of high-energy particles, and also the relative
brightness of various components, namely, of the core, jets, hot spots and
extended components, play an important role in the formation of its integral
radio spectrum~\cite{hardcastle1}. An important issue in understanding the
formation of a radio spectrum is an estimation of the contribution
of physical components of the source to the total radiation in different
frequency ranges and for different orders of radio luminosity.

Multifrequency studies of spectral characteristics of the ``Cold'' survey
sources were repeatedly carried out, starting with the publication of
the RC catalog~\cite{pa2,pa3} itself in 1991--1993, followed by the studies
of 1989--1996 by Bursova et al.~\cite{bur1,bur2,bur3} and of 2001--2006
by Soboleva et al.~\cite{so1,so2}. The spectra of sources of the refined
RC catalog, or the RCR (RATAN Cold Refined)~\cite{so3}, for
the right-ascension interval \mbox{$7^h<RA<17^h$} were also investigated
involving the VLSSr survey~\cite{cohen,lane}, including the flux
density estimates from the VLSSr and GB6~\cite{gregory} survey maps.

With the advent of new surveys in the low-frequency spectral region in
the range of 72--231\,MHz, GLEAM~\cite{hurley} and TGSS~\cite{intema}
with the detection threshold of 50\,mJy and angular resolution from
$25\arcsec$, it became possible to refine the spectra of the RCR sources,
especially for that half of the catalog objects in which the flux data were
known only at two frequencies, 1.4 and 3.94\,GHz. For the interval of
\mbox{$2^h<RA<7^h$} this was already done by Zhelenkova et al.~\cite{zhe1}.
Here we have carried out this task for the entire right-ascension interval
of \mbox{$7^h<RA<17^h$}.

\section{RCR-sources in the interval $2^{h}<RA<7^{h}$ at the wavelength of 7.6\,cm}
The study of Soboleva et al.~\cite{so3}, publishing the RCR catalog
(RATAN Cold Refined), which includes sources from the interval of
\mbox{$07^h<RA<17^h$}, gives the characteristics of objects that are
detected on the scans when they are processed by two methods that differ
the way the background of the sky is determined on the averaged scans.
Reduction of the sky background, as it is done in the $fgr$~\cite{vo} package,
used in the processing, consists of an iterative approximation to a certain
background curve owing to a convolution with the weight function. Varying
the size of the background calculation window, the number of iterations and
the level, we can change the shape of the calculated curve. A rectangle was
used as the weight function, i.e. the sliding averaging was performed.
The first reduction method determined the background by a convolution
with a window of $80^s$, and in the second case, with a window of $20^s$.

Using two different ways of accounting for the sky background, as it turned
out, allows to obtain more accurate results in determining the right
ascension and flux density of the sources. This is also useful for
a more complete source detection. So, for example, three quarters of the
sources were identified in the records using both methods, while
the fourth part of the sources was detected on the scans applying
the first or the second method of sky background approximation.

The reduction results for the right-ascension range of \mbox{$02^h<RA<07^h$},
where the sky background is ``smoothed'' by the $20^s$ window are published
in the study of Zhelenkova et al.~\cite{zhe1}. Here we give the
right ascensions, integral flux densities, and the corresponding
errors of determination of these variables for the list of sources found
on the scans, ``smoothed'' by the window of $80^s$. Thereby, the refinement
of the RC catalog object parameters for this interval of right
ascensions is now complete.

When processing the observational data at the wavelength of 7.6\,cm,
we used standard reduction methods. The reduction procedure is described in
detail in the study of Soboleva et al.~\cite{so3}. The convolution
of averaged scans was performed with the calculated beam pattern in its
central section. Before the identification of objects, the background was
reduced with the window of $80^s$. Next, the sources on the scans were
identified using the gaussian analysis.
The time referencing was done to the strong sources of the NVSS survey~\cite{nvss}.
For each detected radio source its position on the scan $RA$, antenna
temperature $T_a$ and the Half-Power Beam Width (HPBW) were determined.
The integral flux density $F$ was calculated using the formula:
\mbox{$F =k_{eff}\times k_{i}\times T_{a}/k_{DN}$},
where $k_{eff}$ is a coefficient that takes into account the effective area
of the antenna, $k_{i}$  is a correction factor taking into account
the difference of calibrations and a slight difference in the effective area
of the antenna in the observational cycles, $i$ is the number of the cycle,
$k_{DN}$ is a diagram coefficient indicating how much the response from
the source is weakened when it is removed from the central section
of the beam pattern.
The $k_{i}$ coefficient was determined from the sources with well-known
spectra, its value lies in the range of 1.1--1.5 depending on the year
of observations. The value of $k_{eff}$  was equal to 3.5.

The diagram coefficient $k_{DN}$ was calculated for each radio source
with the allowance for a transverse offset of the feed horn along
the focal line of the secondary mirror and the distance of the radio
source along the vertical to the central section by the algorithms presented
in the study of Majorova~\cite{maj1}.

After determining the flux densities and right ascensions of the objects
in each cycle of observations, the mean $F$ and $RA$ over all cycles were
calculated with the corresponding errors.

\subsection{Radio Source List}

Soboleva et al.~\cite{so3} published a summary of sources in the interval
of $07^h<RA<17^h$ and listed the characteristics of the objects found on
the scans, where the background of the sky is ``smoothed'' both by the window
of $80^s$, and $20^s$. The difference in data reduction is called
the first and the second methods, respectively. In this paper, we will adhere
to the format of the table from~\cite{so3} and to the already published
characteristics of the sources~\cite{zhe1}, which are found on the scans,
``smoothed'' by the window of $20^s$, and add the parameters of the sources,
which are identified on the scans, where the background was determined
using a filter with the window of $80^s$.

The sources discovered by both the first and second methods in the interval
\mbox{$2^h<RA<7^h$} are presented in Table~\ref{Tab1}, where column (1)
lists the source name, composed of the source coordinates $RA_{2000}$ and
$Dec_{2000}$ from the NVSS~\cite{nvss} catalog.
Columns (2) and (3) contain the right-ascension differences
of the objects $\Delta RA_{1}$ and $\Delta RA_{2}$ with errors, where
\mbox{$\Delta RA_{1}=RA_{NVSS}-RA_{1}$} and
\mbox{$\Delta RA_{2}=RA_{NVSS}-RA_{2}$}, $RA_{NVSS}$ is the right ascension
of the source according to the NVSS data.
Columns (4) and (5) list the flux densities of the sources $F_{1}$ and
$F_{2}$ in mJy with the inaccuracies (as the average over all
the observational cycles).
$RA_{1}$ and $F_{1}$, $RA_{2}$ and $F_{2}$ are respectively
the right ascensions and flux densities, obtained during the processing of
observational data by E.\,K.~Majorova using the first method, and by
 N.\,S.~Soboleva with A.\,V.~Temirova using the second method (see Table 1
in~\cite{zhe1}).\\
Columns (6) and (7) give the values of spectral indices $\alpha_{3.94}$
and $\alpha_{0.5}$ \mbox{($F_{\nu}\propto\nu^{\alpha}$)}, which are
determined at the frequencies of 3.94 and 0.5\,GHz, respectively.
The latter frequency was chosen by analogy with the study of Miley and
De Breuck~\cite{mi}.\\
Column (8) ``Mrph'' provides comments on the morphology of the sources,
determined from the NVSS and TGSS survey maps, as well as the features of
registration of sources by the RATAN-600 beam pattern:\\
-- ``d'', ``m'' is a source with two or more NVSS-components respectively;\\
-- ``b'' is a blend, i.e. two or more sources are registered as one due to
the shape and size of the telescope's beam pattern;\\
-- ``R'' is a blended source that can be resolved by software.\\
Column (9) ``Flx'' lists comments on the features of flux density
measurements and variability. The last based on the literature or the simple
condition $F_{var}>3$, where
\mbox{$F_{var}=(F_{max}-F_{min})/\sqrt{(err_{max}^2+err_{min}^2)}$}:\\
-- ``V'' is a variable, according to the literature, radio source, or ``v'',
suspected of variability~\cite{maj2};\\
-- ``F'' -- the source flux density in the FIRST survey is higher
($F_{var}>3$) than that in the NVSS, which is a sign of possible variability;\\
-- ``B'' -- there is a significant ($F_{var}>3$) flux density scatter at
the frequencies of 3.94--5\,GHz according to the catalog data;\\
-- ``E'', ``e'' -- the estimates of the source flux density from the GB6 survey
maps are brighter ($F_{var}>3$) or, respectively, weaker than ought to be
from a comparison of the estimated flux density and the one measured at
the frequency of 3.94\,GHz;\\
-- ``s'' (scattered) is a large spread of the flux density data at different
frequencies;\\
-- ``\#'' -- the flux density data is available only at the
frequencies of 1.4 and 3.94 GHz;\\
-- ``1'' -- detected only in one cycle of ``Cold'' experiment.\\
Column (10) ``Sp.'' lists the types and features of the radio spectra:\\
-- ``l'' is the spectrum where the flux density $F$ dependence on $\nu$ is
described by a power function ($F\propto\nu^{\alpha}$) and is approximated
on the logarithmic scale by a straight line (S-spectrum);\\
-- ``+'' and ``-'' -- the spectrum at higher frequencies gets flatter (C+)
or steeper (C-), respectively;\\
-- ``h'' (hill) -- the complex spectrum is formed by imposing a power-law
spectrum on a spectrum with self-absorption at the frequencies
from 0.1 to 12\,GHz. In this case the spectral index $\alpha_{0.5}$
was determined from the spectrum built from the data up to 1\,GHz,
while $\alpha_{3.94}$ -- from 1\,GHz and higher;\\
-- ``g'', ``p'' -- is the spectrum with a maximum in the range of GHz or MHz,
respectively;\\
-- ``G'' -- is a well-known source of GPS (GigaHertz Peak Spectrum);\\
-- ``u'' (upturn) -- there is a minimum in the spectrum, followed by growth
at the frequencies above 5\,GHz.

We recall that the altitude of observations for the RATAN-600 antenna in
the surveys was chosen by the visible coordinates of the SS 433 object
from the beginning of the ``Cold'' experiment in 1980 and was retained in
the subsequent cycles, i.e. the investigated strip of the sky was slightly
offset in the declination from cycle to cycle. Thus, some sources could be
registered only in one survey~\cite{zhe2}. This refers to 065848.74+045522.0.
All the other sources in the published list were found in at least two cycles
of the ``Cold'' experiment.
\begin{figure*}[]
\setcaptionmargin{5mm}
\onelinecaptionsfalse
\centerline{
\hbox{
\includegraphics[angle=0,width=0.5\textwidth,clip]{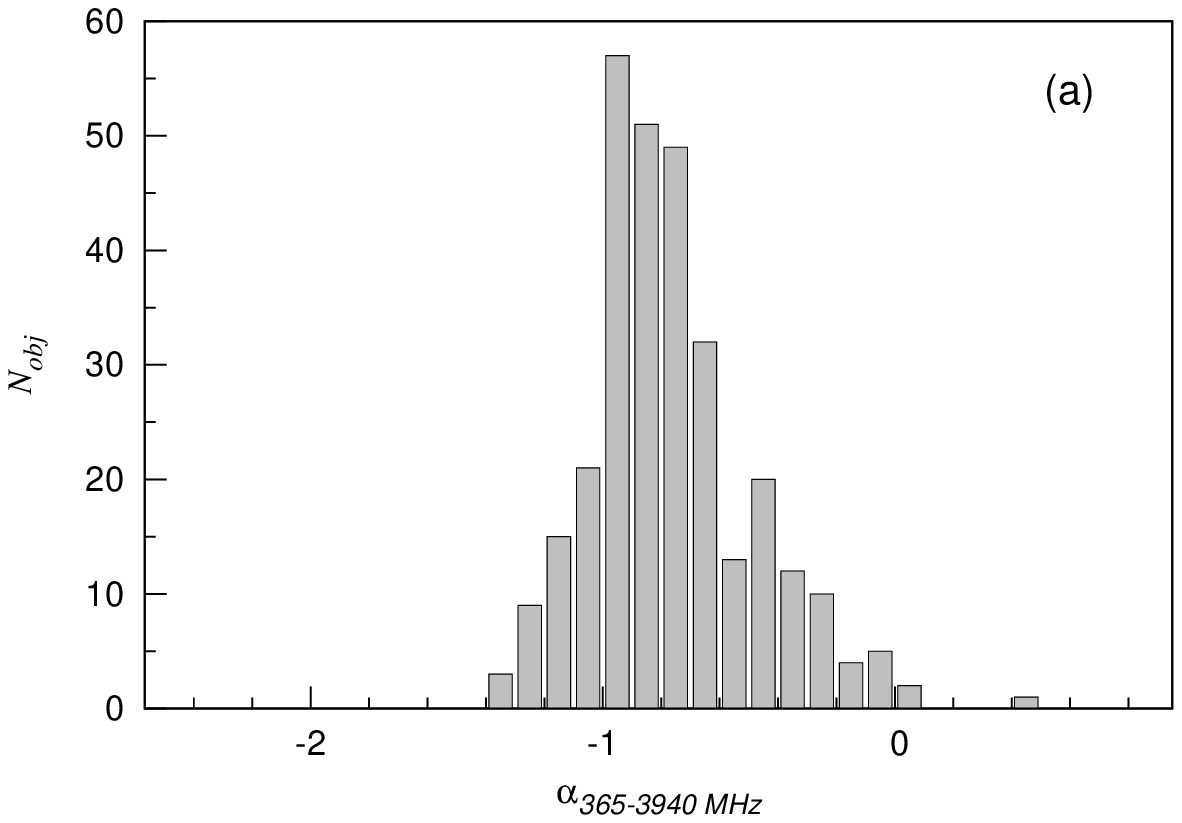}
\includegraphics[angle=0,width=0.5\textwidth,clip]{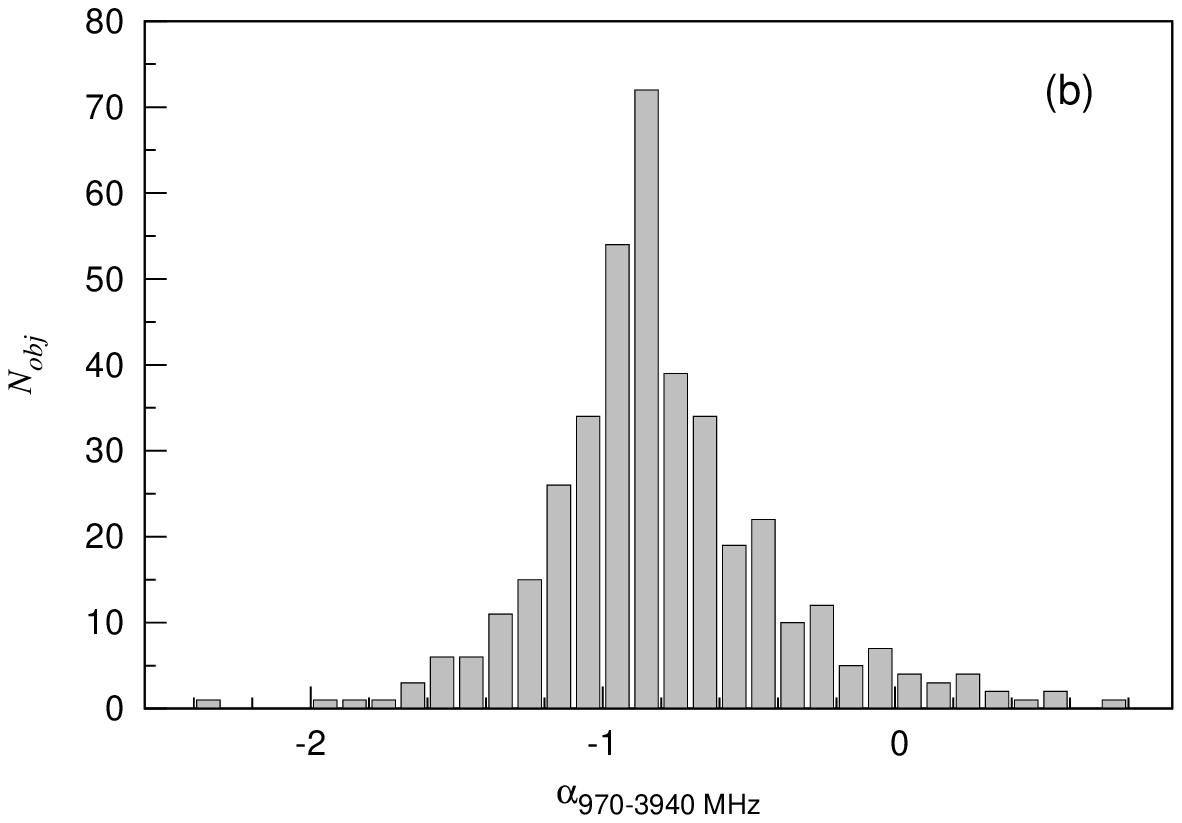}
}
}
\caption{
Histograms with the distribution of spectral indices for the RC catalog radio
sources: (a) for 304 objects, for which the spectral indices were determined
in the papers~\cite{pa2,pa3}, at the frequencies of 365--3940\,MHz;
(b) for 396 sources from~\cite{bur3} at the frequencies of 970--3940\,MHz.
}
\label{Fig1}
\end{figure*}

A total of 256 sources were identified in the right-ascension interval
$2^h<RA<7^h$. Of these, 68\% are found using two methods and 32\% -- applying
the first or the second method.

\section{SPECTRAL CHARACTERISTICS OF RCR SOURCES
}

In the literature, publications on the multifrequency studies of the spectra
of radio sources are not so common. As an example, one can cite one of the
first papers of Laing and Peacock~\cite{laing}, and then:
Ker et al.~\cite{ker}, Mahony et al.~\cite{mahony},
Whittam et al.~\cite{whittam}, Calistro Riviera et al.~\cite{calistro}.

Laing and Peacock~\cite{whittam} found in the 178--2700\,MHz frequency range
the dependence of the shape of the radio source spectra on luminosity. More
powerful sources have a flatter spectrum, while weaker sources reveal
a steepening of the spectrum at low frequencies.

Wittam et al.~\cite{whittam} noted the enhancement of the flux for
radio galaxies at high frequencies (15.7\,GHz) and suggested that this could
be due to the nuclei of the FRI-type sources that become dominant in this
range.

Investigations of a sample of radio galaxies from the LOFAR deep survey in
the range from 150\,MHz (the limit of 120--150$\mu$Jy) and up to 1.4\,GHz
(28$\mu$Jy) have shown that the steepening at low frequencies, observed for
faint AGNs can be explained by the domination of radio source components with
a steep spectrum, although the contribution of components with a flat spectrum
becomes more significant at high frequencies, making the spectral index more
flat~\cite{calistro}. Due to a rather high uncertainty in the flux density
estimation owing to different angular resolutions in the surveys and methods
of source detection, this result may be rather ambiguous for AGNs since they
are extended sources.
\begin{figure*}[]
\setcaptionmargin{5mm}
\onelinecaptionsfalse
\centerline{
\hbox{
\includegraphics[angle=0,width=0.5\textwidth,clip]{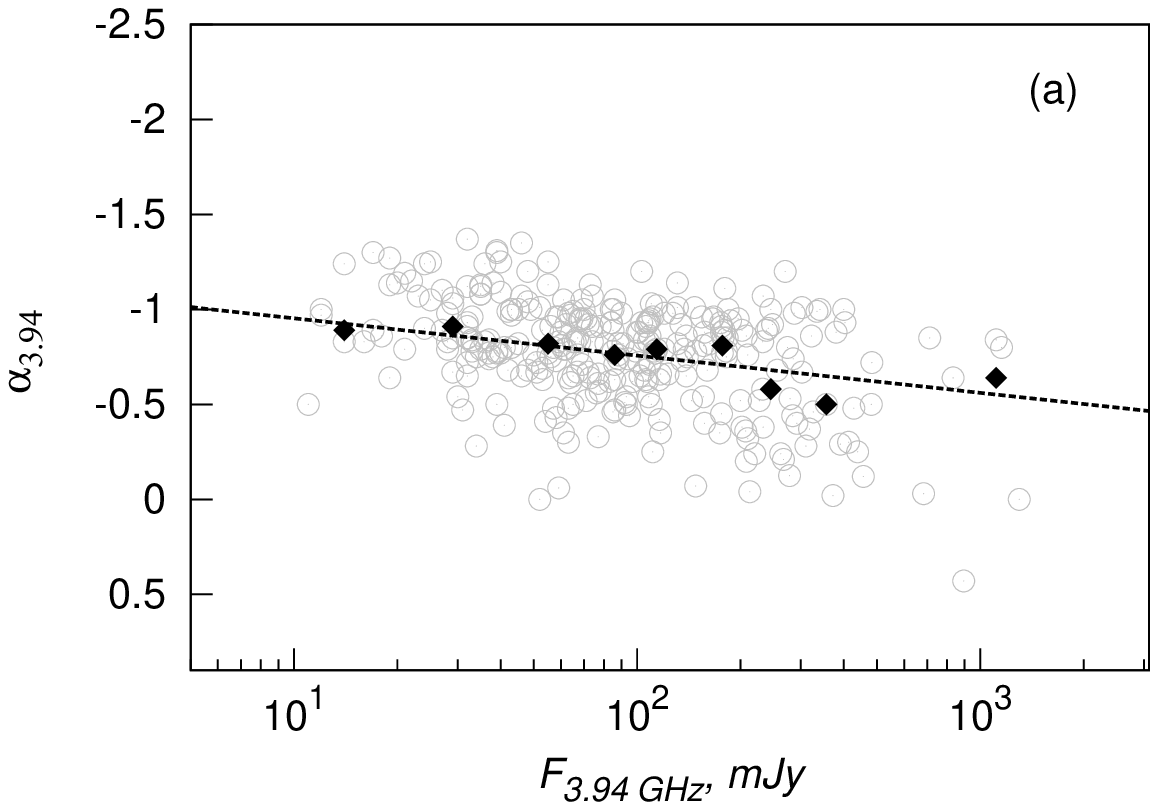}
\includegraphics[angle=0,width=0.5\textwidth,clip]{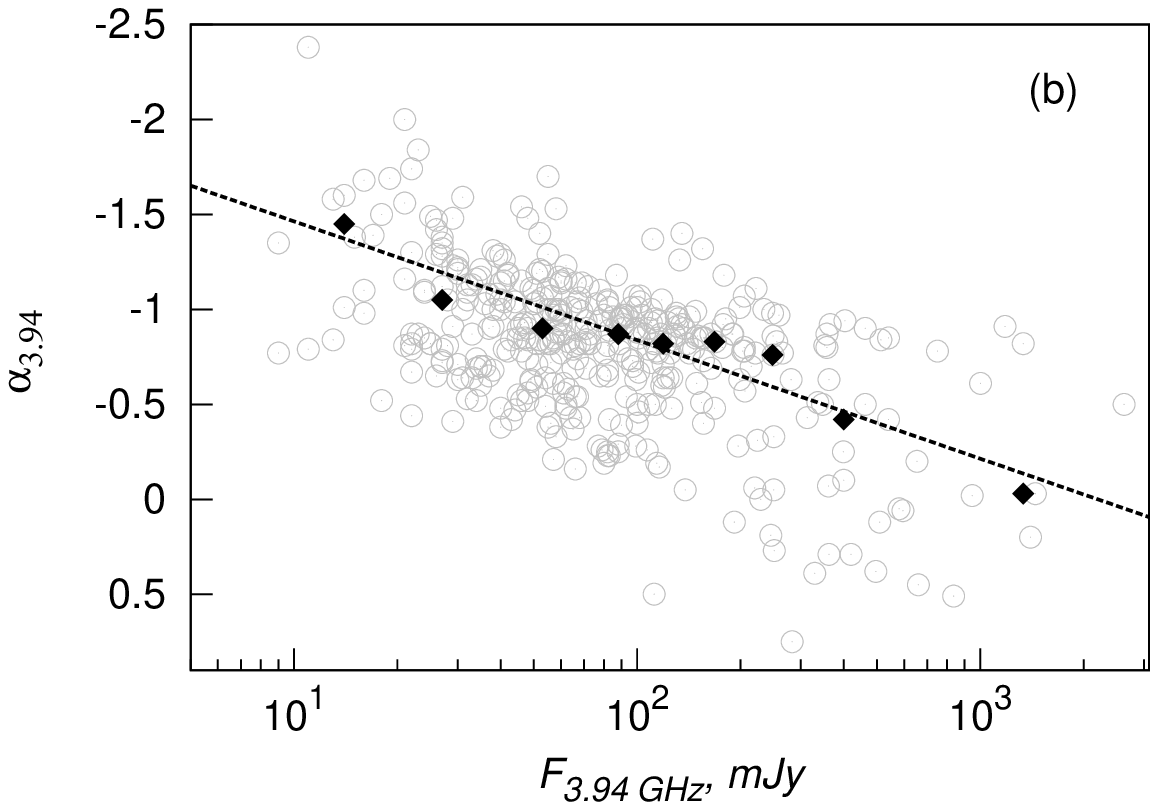}
}
}
\caption{
The scattering diagram for the spectral indices and integral flux density at
the frequency of 3.94\,GHz $F_{3.94}$ for the RC catalog radio sources:
(a) for 304 objects from~\cite{pa2,pa3} and (b) for 396 sources
from~\cite{bur3}. The values for the sources are shown in the figures by
gray circles. Each list was divided into the bins based on the flux density.
The median value of the spectral index and the flux density of each bin is
shown on the plot with a black square. The dashed line describes the linear
regression.
}
\label{Fig2}
\end{figure*}

The spectral characteristics of the sources that were detected in
the ``Cold'' experiment surveys have been repeatedly determined both from
the multifrequency observations conducted at the RATAN-600, and with
the involvement of data from well-known catalogs.

The sensitivity limit of the ``Cold'' experiment surveys in 1980--1981
was significantly better than the ones of all the catalogs available then.
Therefore, in the RC catalog publication~\cite{pa2,pa3} the spectral indices
in the frequency range of 365--3940\,MHz were determined only for one
fourth of the sources. Rather bright sources made it into this sample.
The median values of the flux density and spectral index for it amounted
to $F_{3.94}=82\,mJy$ and $\alpha=-0.82$, respectively. Figure~\ref{Fig1}a
presents a histogram with the distribution of spectral indices from
the materials of the papers~\cite{pa2,pa3}.

In one of the first publications on the spectral characteristics of
RC sources~\cite{bur1}, the data was mainly obtained from the multi-frequency
observations at the RATAN-600. The spectra were constructed for one
third of the 840 objects from the first list of the RC catalog~\cite{pa2}.
In 70\% of them, the spectra turned out to be steep with an average spectral
index of $\alpha=-0.87$.

In the studies of Bursov et al.~\cite{bur2,bur3} the spectra for 529 (44\%)
RC sources in the range of 970--3940\,MHz were mainly determined from
the observations at the RATAN-600 radio telescope. Most of them are bright
enough objects ($F_{3.94}$>35\,mJy) with steep spectra at the median values
of $F_{3.94}$=69\,mJy and $\alpha$=--0.85. There is an insignificant number
of sources (3\%) with inverse spectra ($\alpha\ge0.1$). Figure~\ref{Fig1}b
presents a histogram with the distribution of indices for the sources
from~\cite{bur3}.

We have divided the sources for which the spectral indices are determined
in the papers~\cite{pa2,pa3,bur3} into groups by the flux density,
taking as the unit of measurement the noise error of $\sigma$=3.5\,mJy as the
average of the root-mean-square noise errors from the 1980--1994
surveys~\cite{so3}. The first bin included the sources with $F\le$10.5\,mJy
at the frequency of 3.94\,GHz, i.e. weaker than $3\sigma$. The boundaries of
the following bins were chosen to be 17.5, 35, 70, 105, 140, 210, 280, 700
and over 700\,mJy or $5\sigma$, $10\sigma$, $20\sigma$, $30\sigma$,
$40\sigma$, $60\sigma$, $80\sigma$, $200\sigma$  and over $200\sigma$,
respectively. Then we slightly changed the division for an even more
homogeneous distribution of sources into groups, so that in each of them
there would be enough sources for the statistics. As a result, the
upper boundaries of bins were as follows: 15, 25, 35, 50, 70, 100, 150, 250
and more than 250 mJy.

Figure~\ref{Fig2} shows a scattering diagram for the spectral indices and
integral flux density $F_{3.94}$ from the data of~\cite{pa2,pa3}
(Fig.~\ref{Fig2}a) and~\cite{bur3} (Fig.~\ref{Fig2}b). The values for
the sources are shown by gray circles, the median value of the spectral index
and the flux density of each bin -- by a black square, the regression line
from these values -- by a dashed line. Both in the first (Fig.~\ref{Fig2}a)
and in the second case (Fig.~\ref{Fig2}b) rather bright sources, discovered
within the ``Cold'' experiment at 3.94\,GHz reveal a tendency to the
flattening of the spectra with increasing flux density.

In the works of Soboleva et al.~\cite{so2,so3} the NVSS catalog was used
to determine the two-frequency indices $\alpha^{1.4}_{3.94}$ at 1.4
and 3.94\,GHz (the wavelength of 20 and 7.6\,cm, respectively) in RC sources,
including weak objects ($F_{3.94}$<35\,mJy). In the $10'$-wide band
extending for about $16^h$ in the right ascension, 95\% of the RC catalog
objects were identified with the NVSS sources. It was found that with
increasing flux density at the wavelength of 7.6\,cm, spectral indices of the
sources become more steep.

According to the data obtained over 1987--2000, for the ``Cold'' strip
extending for $11^h$ over the right ascension, about 600 sources were
detected~\cite{so3}, which were identified with the NVSS. The average
value of the two-frequency spectral index $\alpha^{1.4}_{3.94}$
for them was --0.44. A half of these sources proved to have a standard
power spectrum, while in 30\% of objects the spectrum becomes steeper with
increasing frequency. The flattening of the two-frequency spectral
index $\alpha^{1.4}_{3.94}$ is observed at weaker flux densities,
as it was also noted in~\cite{so2}, which may be due to a decrease in
the proportion of FRII-type sources.

\subsection{Spectral Indices of RCR Sources and New Data}

With the advent of more flux density-sensitive TGSS and GLEAM surveys,
it became possible to refine the spectra of bright sources in
the low frequency range, and also to more confidently determine
the spectra of a half of the RCR sources for which the data on the flux
density were available only at 1.4 and 3.94\,GHz.

To construct the source spectra, we used the mean values from the integral
flux densities obtained by two methods at 3.94\,GHz, the VLSSr, GLEAM,
TGSS, TXS~\cite{txs}, NVSS, FIRST and GB6 survey data, as well as
the information from the CATS~\cite{v1,v2}, Vizier~\cite{vizier} and
NED databases.
\begin{figure*}[]
\setcaptionmargin{5mm}
\onelinecaptionsfalse
\centerline{
\hbox{
\includegraphics[angle=0,width=0.5\textwidth,clip]{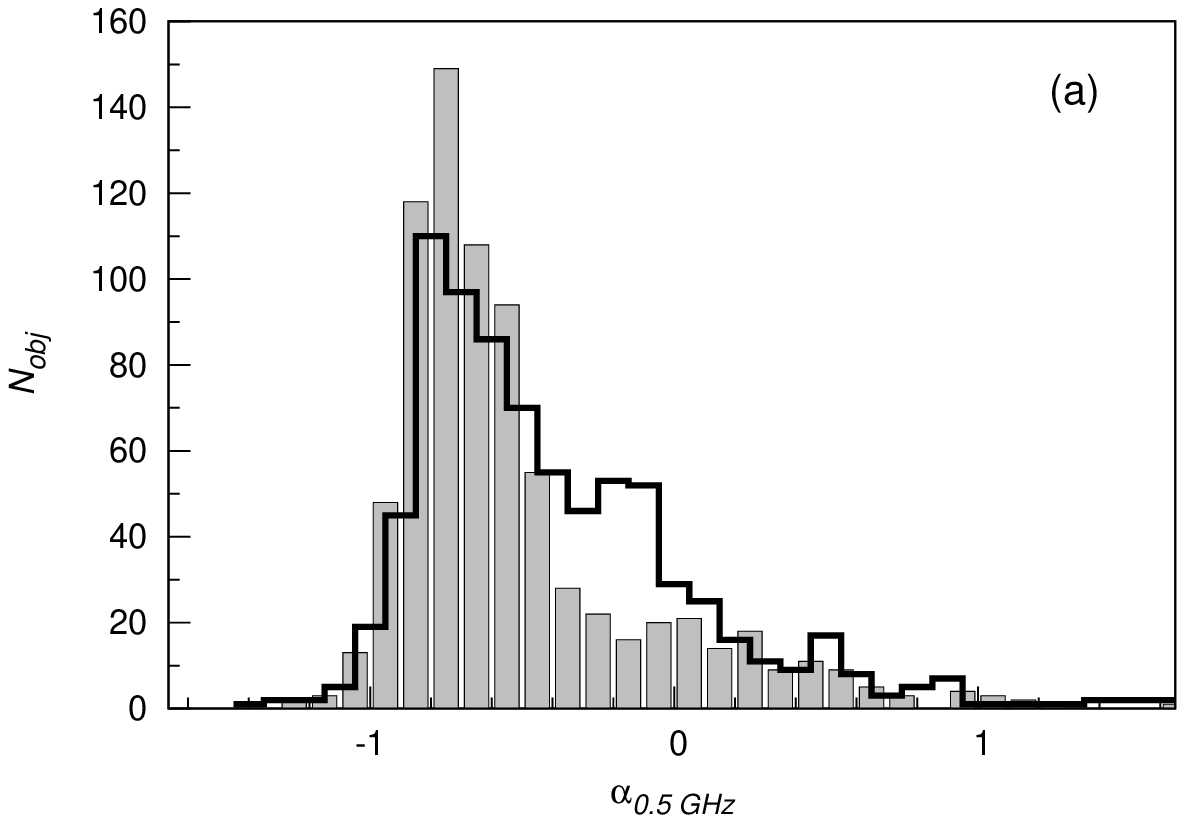}
\includegraphics[angle=0,width=0.5\textwidth,clip]{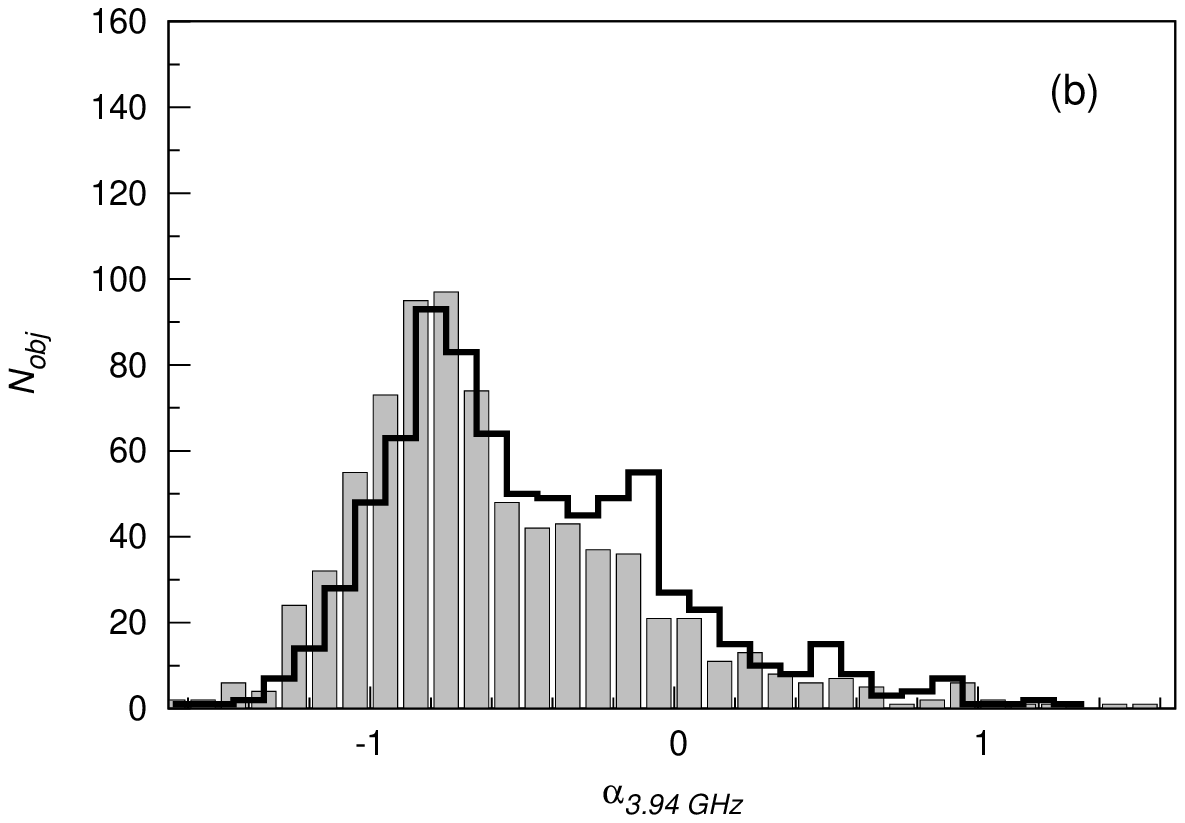}
}
}
\caption{
The distribution of spectral indices of 830 RCR catalog sources for
the right ascension interval (\mbox{$2^{h}< R.A.< 17^{h}$}) at the
frequencies of 0.5\,GHz (a) and 3.94\,GHz (b). The gray bars denote
spectral indices determined from the spectra constructed invoking the data
from the GLEAM and TGSS surveys, the black broken line -- without these data.
}
\label{Fig3}
\end{figure*}

Now there are 13\% of the objects whose flux densities are known at two
frequencies. They are not available in the VLSSr, GLEAM, TGSS and GB6
catalogs, since their flux is weaker than the detection threshold of
reliably identified sources adopted in the catalogs. For a number of such
sources at the visual control, a noticeable brightening was revealed on the
maps. To construct their spectra, we engaged the the flux density estimates
based on the VLSSr (74\,MHz), TGSS (150\,MHz) and GB6 (4.85\,GHz) survey
radio maps, sometimes also using the GLEAM (the band of 70--231\,MHz).
These estimates are performed using the Aladin
functions~\cite{aladin1, aladin2}. To work with the tables, we used
the TOPCAT~\cite{topcat} software application.

Applying the software package $spg$~\cite{vo} the approximation of spectra
was done by the polynomials of the first or second degree, and spectral
indices $\alpha_{3.94}$ and $\alpha_{0.5}$  were calculated at
the frequencies of 3.94 and 0.5\,GHz, respectively, for the list sources.

Let us compare the way how the distribution of spectral indices of
the sources found in the ``Cold'' surveys varied as the new data appeared in
the other frequency ranges.

In their publication, Soboleva et al.~\cite{so1} engaged the VLSS and GB6
surveys for constructing the spectra of the RCR catalog sources. New low
frequency GLEAM and TGSS surveys were used at the next iteration for
the construction of spectra of RCR-sources from the interval of
\mbox{$2^{h}<R.A.<7^{h}$}~\cite{zhe1}. Continuing it here, we present
the results for all the RCR sources from the interval of
\mbox{$2^{h}<R.A.<17^{h}$} and compare the spectral indices computed from the
spectra, where the GLEAM and TGSS survey data were used in the approximation
(further referred to as ``new'' here), and the indices obtained without the
GLEAM and TGSS data (``old'').

Figure~\ref{Fig3} presents the distributions of spectral indices of 830
RCR-sources at the frequencies of 0.5 and 3.94\,GHz. The ``new'' spectral
indices are shown by gray bars, and the ``old'' ones -- by the broken black
line. The histograms reveal a noticeable difference for the ``old'' and
``new'' indices, which can be associated with the refinement of radio spectra
for those sources whose flux data had previously been only known at two
frequencies, and the spectrum was usually approximated by a straight line.
Thus, the number of spectra that are best approximated by a straight line
(the S-spectra)\footnote{
We compared the spectral indices at 0.5 and 3.94 GHz.}
decreased from 73\% (``old'' spectral indices) to 35\% (``new'' spectral
indices).

Dividing the sources into groups with ultra steep (USS, Ultra Steep Spectrum;
\mbox{$\alpha\le-1.0$)}, steep (SS, Steep Spectrum;
\mbox{$-1.0<\alpha\le-0.5$)}, flat (FS, Flat Spectrum;
\mbox{$-0.5<\alpha\le0.1$)} and inverted (IS, Inverted Spectrum;
$0.1<\alpha$) spectra, we have obtained the following composition at
the frequency of 0.5\,GHz (Fig.~\ref{Fig3}a) with the median value of
the spectral index for each group:
USS -- 2\%, $\alpha=-1.02$;
SS -- 68\%, $\alpha=-0.74$;
FS -- 21\%, $\alpha=-0.32$;
IS -- 9\%, $\alpha=0.38$,
and at the frequency of 3.94\,GHz (Fig.~\ref{Fig3}b):
USS -- 14\%, $\alpha=-1.12$;
SS -- 52\%, $\alpha=-0.77$;
FS -- 24\%, $\alpha=-0.28$;
IS -- 10\%, $\alpha=0.42$.

The statistics for other types of spectra has also changed with
the addition of GLEAM and TGSS survey data.
In 10\% of the sources, the maximum of the spectrum falls on
the frequencies of 0.1--12\,GHz. Such spectra are observed in  CSS
(Compact Steep Spectrum), GPS (Gigahertz Peak Spectrum), HFP (High Frequency
Peaker). Seven objects from them are included in the published lists
of such sources.
And still 3\% of sources have a maximum in the spectrum, but the spectrum
has yet a more complex shape. It is formed by imposing a power-law
spectrum at low frequencies onto a spectrum with self-absorption at
the frequencies of 0.5--12\,GHz. Sources with such a radio spectrum are
indicated in Table~\ref{Tab1} with the symbol ``h''.
In our list there are 3\% of sources with upturn-spectra, where the spectra
reveal a minimum, followed by a growth at the frequencies from several
gigahertz.

\subsection{Spectral Indices and Flux Density}

Samples which are complete with regards to the flux density of radio sources
can include different populations of objects, depending on which frequency
and with what limit of flux density they are obtained, since due to different
redshifts of objects, the contributions to the integral observable spectrum
come from various physical components of the source (see~\cite{ker}, Fig. 2).
If the source is observed not along the axis of the jet, then at
low frequencies synchrotron radiation from extended components shall dominate.
At the frequencies above several gigahertz, the contribution to radiation
from the nucleus becomes significant.
\begin{figure*}[]
\setcaptionmargin{5mm}
\onelinecaptionsfalse
\centerline{
\vbox{
\hbox{
\includegraphics[angle=0,width=0.5\textwidth,clip]{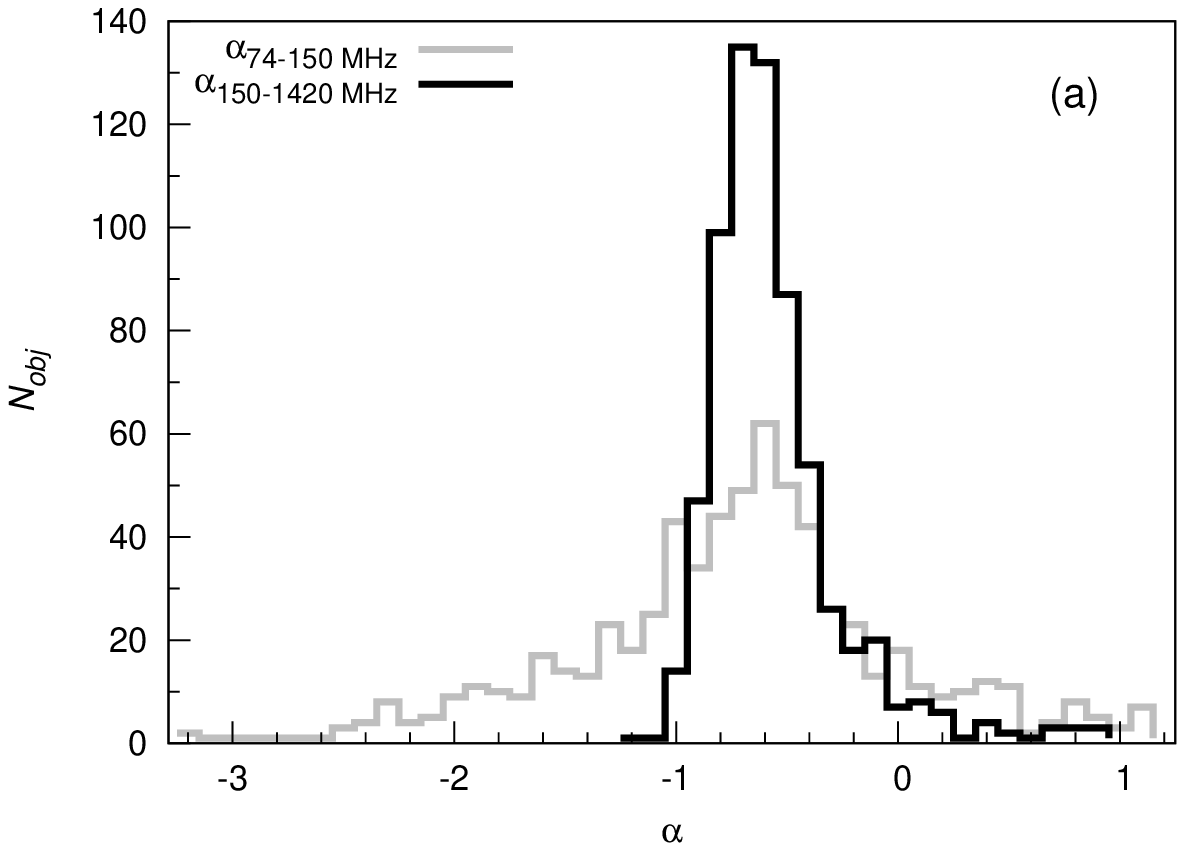}
\includegraphics[angle=0,width=0.5\textwidth,clip]{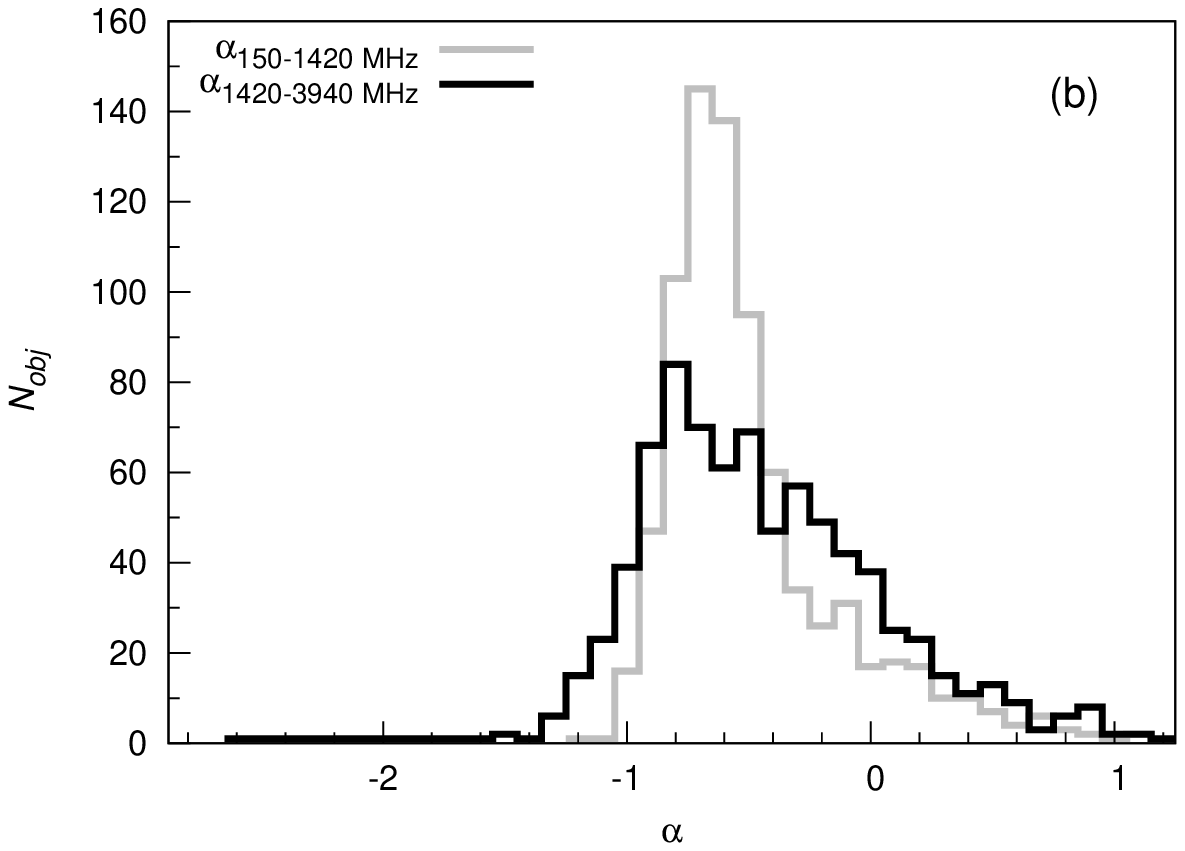}
}
\hbox{
\includegraphics[angle=0,width=0.5\textwidth,clip]{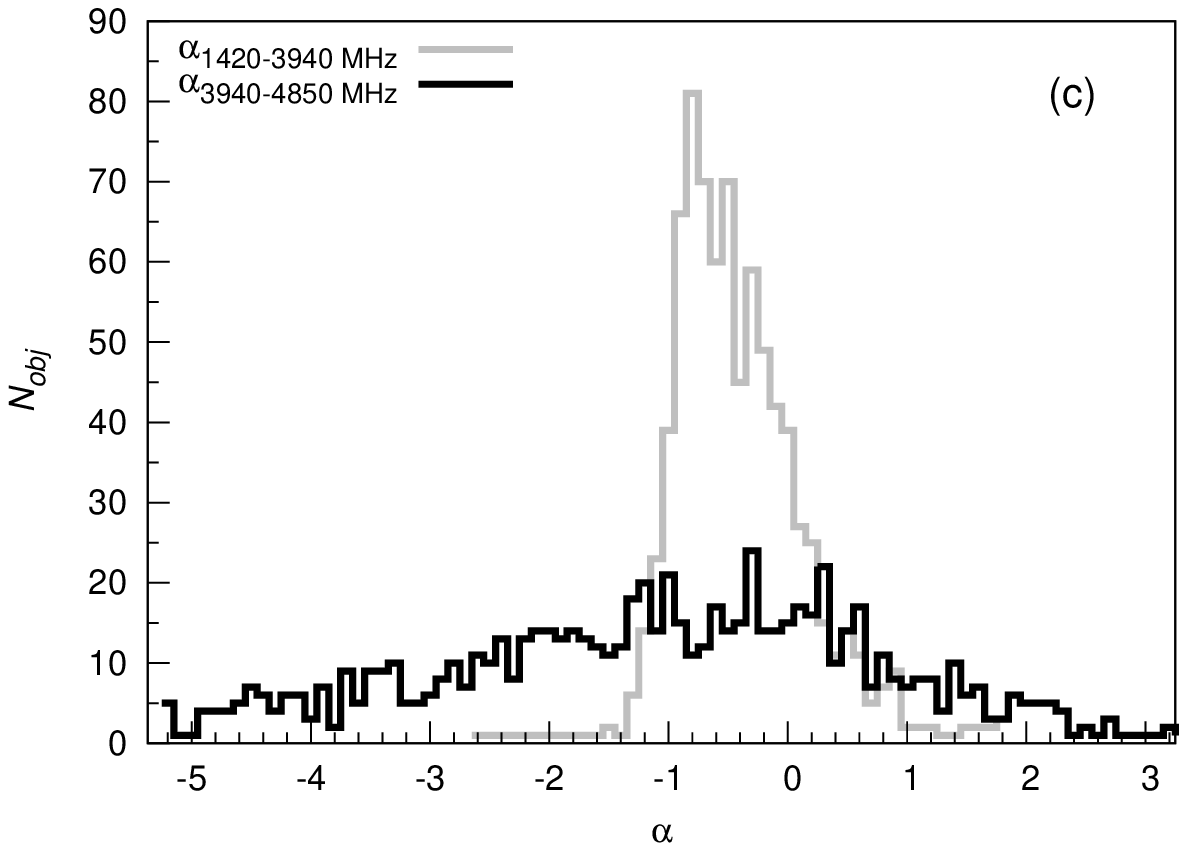}
\includegraphics[angle=0,width=0.5\textwidth,clip]{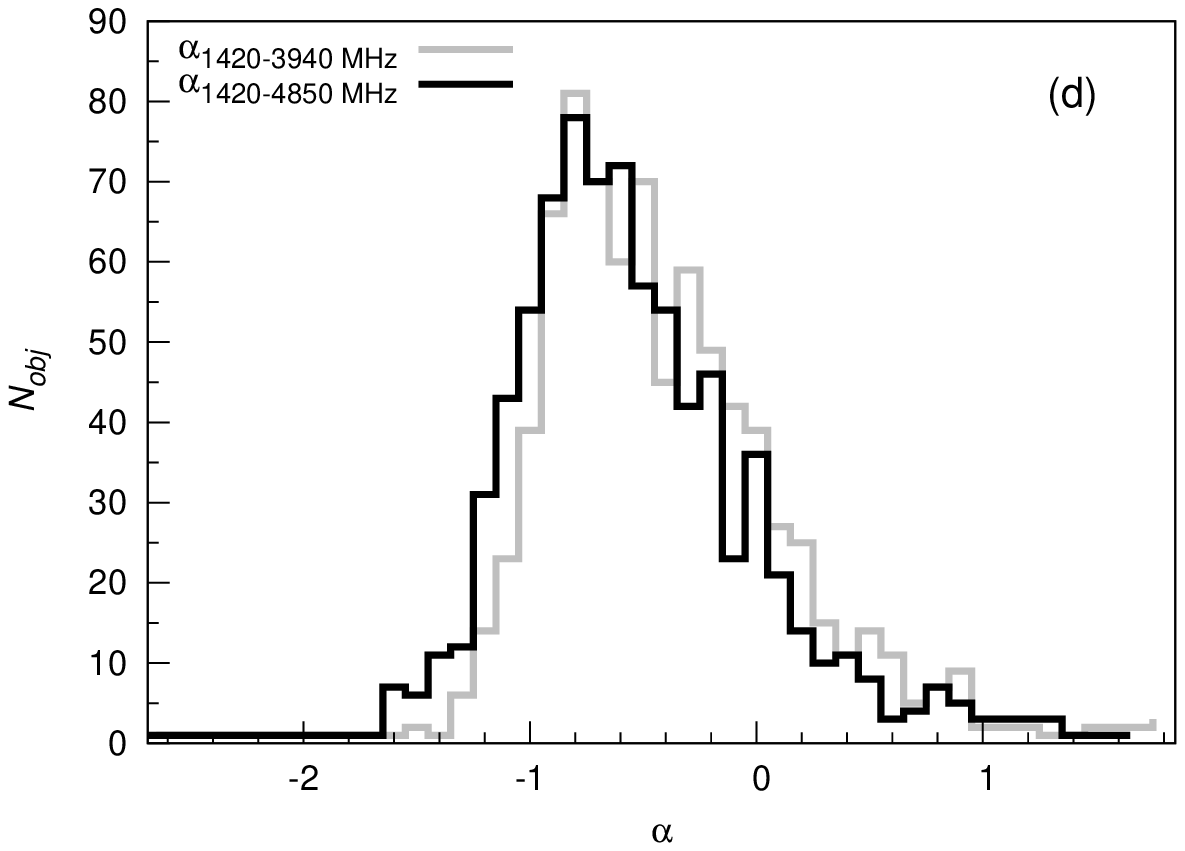}
}
}
}
\caption{
The distribution of two-frequency spectral indices for RCR catalog sources.
The diagram (a) shows in gray the indices at the frequencies of
74--150\,MHz ($\alpha_{74-150}$), the black broken line -- the indices at
the frequencies of 150--1420\,MHz ($\alpha_{150-1420}$);
(b) the two-frequency indices $\alpha_{150-1420}$ and $\alpha_{1420-3940}$;
(c) $\alpha_{1420-3940}$ and $\alpha_{3940-4850}$;
(d) $\alpha_{1420-3940}$ and $\alpha_{1420-4850}$.
}
\label{Fig4}
\end{figure*}
If the sampling is performed at 100--300\,MHz, then up to large redshifts
($4\leq z\leq 6$) the radio emission will still be registered from
the extended components, which allows to directly compare the characteristics.
If the sampling is done at the frequency of 1\,GHz or higher, then the sources
with a significant contribution of the nucleus, which are oriented so
that the jet is directed along the line of sight, would more preferably get
into the sample, and the more often, the greater their redshift. Note that
most of the sources known at $z>4$ are compact and often reveal a bent or
a peaked spectrum, what indicates the youth of the source or a certain
activity, related with the merger of galaxies. Since our sample is produced at
the frequency of about 4\,GHz, then it should have a significant share
of sources with a substantial contribution of the nucleus into
the integral radio spectrum.

Let us consider at the histograms (Fig.~\ref{Fig4}) the way the distributions
of two-frequency spectral indices for the RCR-sources vary in the frequency
range from 74\,MHz to 4.85\,GHz. Each diagram in Figure~\ref{Fig4} presents
two histograms with the distribution of two-frequency indices: for lower
frequencies the histogram is represented by the gray bars, and for
higher frequencies -- with a black broken line. The most compact distribution
of indices was obtained for a two-frequency spectral index
$\alpha^{0.15}_{1.4}$, and the largest scatter -- for $\alpha^{3.94}_{4.85}$.

Comparing these diagrams we may assume variations in the contribution of
different physical components of a radio source into the integral spectrum.
Thus, at 74--150\,MHz synchrotron self-absorption in the shock region is
most likely at work. For the frequency range of 0.15--1.4\,GHz, the main
contribution is provided by the synchrotron radiation of extended components,
while the physical conditions in them are apparently quite close.
At the frequencies of 1.4--3.94\,GHz, the radiation of the nucleus is
added, and at the frequencies of 3.94--4.85\,GHz its contribution becomes
more significant. On top of this, we have to account both for the different
power of the sources and their evolutionary stage -- the initial phase,
where the jet starts to break through the environment, and then a formed jet,
where its infeed by the nucleus still carries on, and the relic phase, where
the infeed of the jet has already ceased.
\begin{figure*}[]
\setcaptionmargin{5mm}
\onelinecaptionsfalse
\centerline{
\vbox{
\hbox{
\includegraphics[angle=0,width=0.5\textwidth,clip]{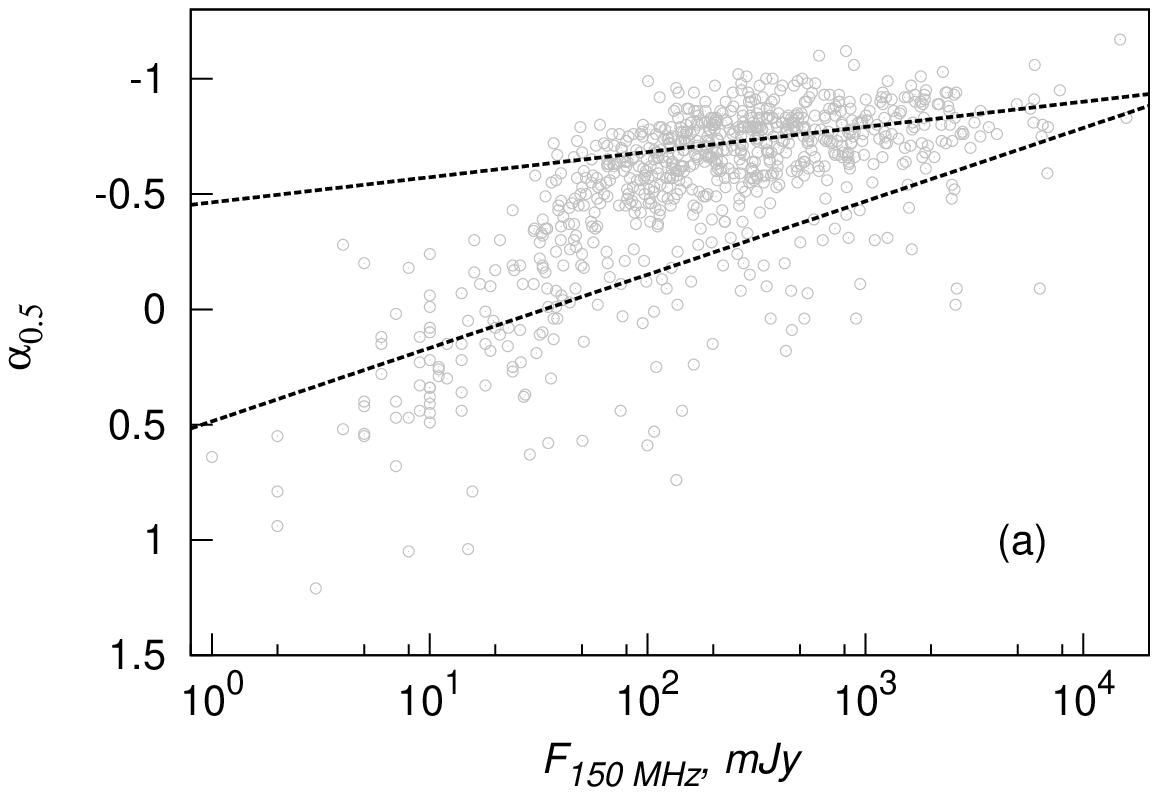}
\includegraphics[angle=0,width=0.5\textwidth,clip]{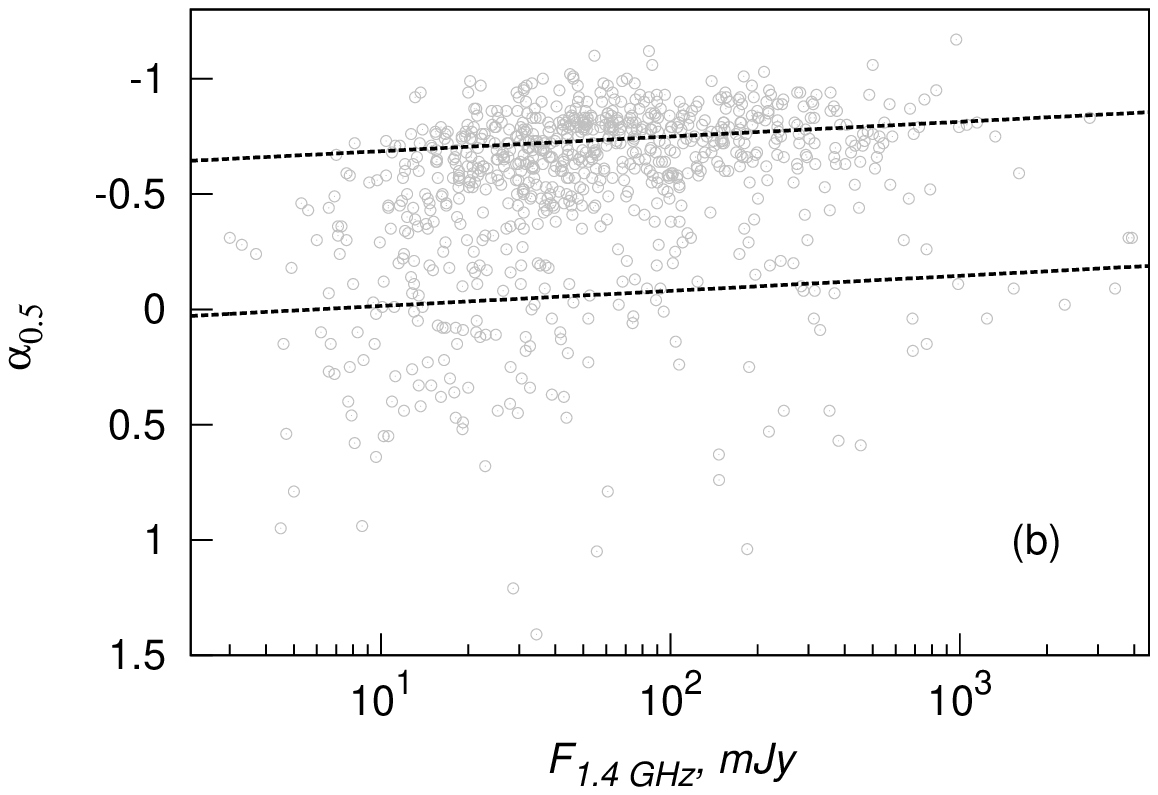}
}
\hbox{
\includegraphics[angle=0,width=0.5\textwidth,clip]{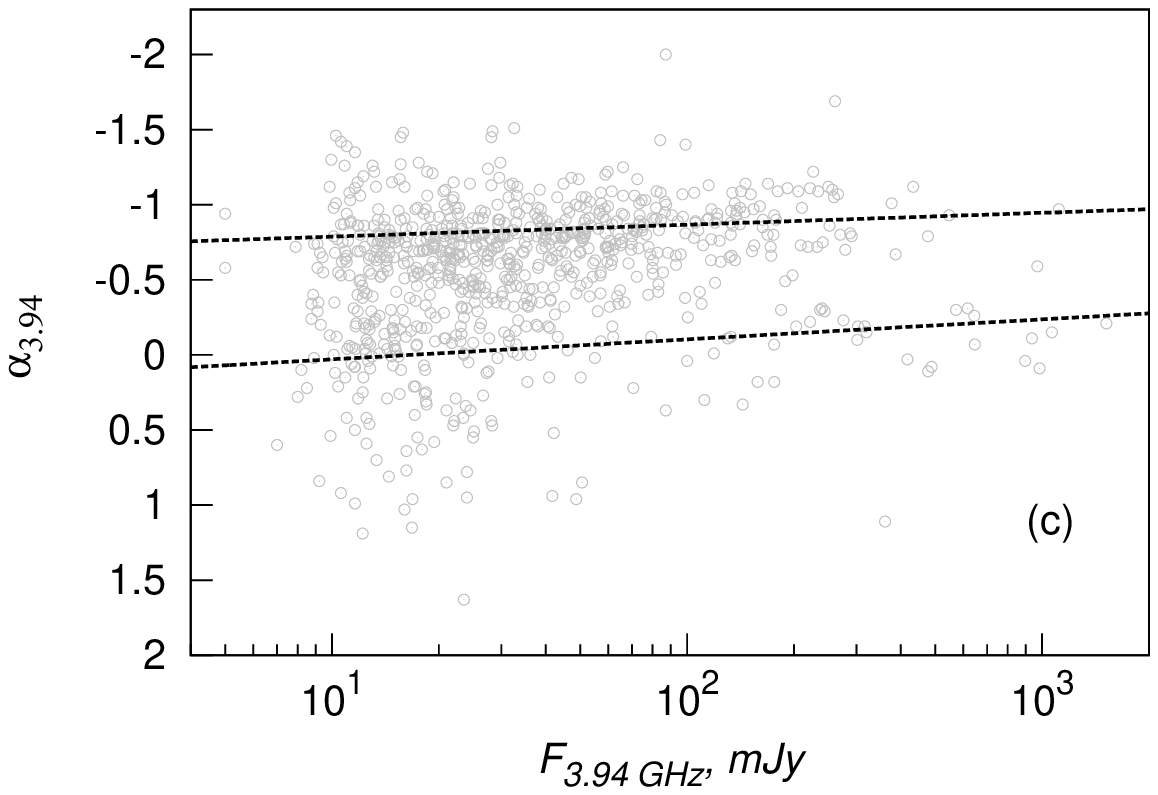}
\includegraphics[angle=0,width=0.5\textwidth,clip]{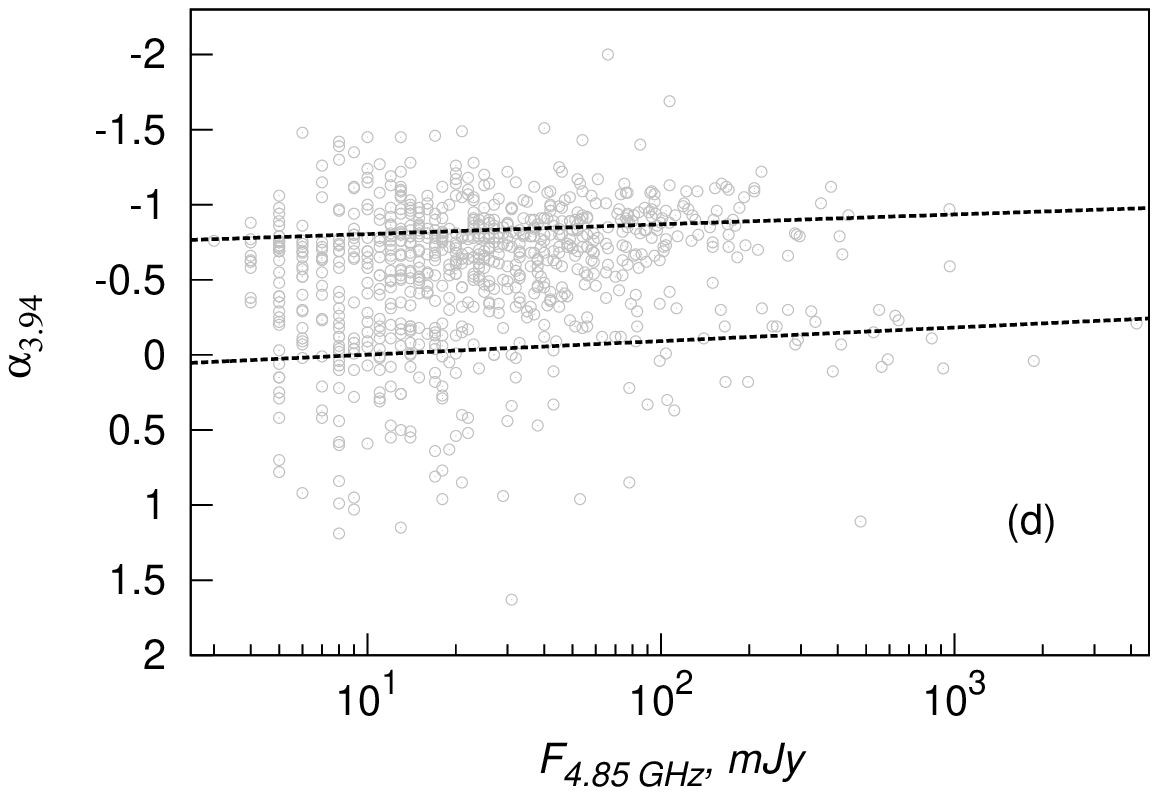}
}
}
}
\caption{
The diagram of scattering of spectral indices at the frequencies of
0.5\,GHz (a, b), 3.94\,GHz (c, d) and the flux density at the frequencies of
150\,MHz (a), 1.4 (b), 3.94 (c) and 4.85\,GHz (d) for 830 sources of
the RCR catalog. The upper dashed line is a regression line over all points
for the sources with steep spectra ($\alpha\le-0.5$), bottom dashed line is
for the sources with flat spectra.
}
\label{Fig5}
\end{figure*}

To determine the possible relationship between the spectral index and
the integral flux density, which was indicated in~\cite{so2,so3} for
the RC-catalog sources, we have considered various combinations of spectral
indices and flux densities at different frequencies for 830 RCR-catalog
sources (Fig.~\ref{Fig5}). On the diagrams, linear regression across all
the points for the sources with steep spectra, SS ($\alpha\le-0.5$)
is described by the upper gray dashed line, and for the sources with
flat spectra, FS, ($\alpha>-0.5$) -- by the bottom dashed line. Note that
the SS-sources account for 70\%, if we apply $\alpha_{0.5}$ for the division
into groups, and 66\%, if we use $\alpha_{3.94}$.

There is a larger concentration of sources with steep spectra in a certain
area of the diagram in all figures as compared to the sources with flat
spectra, which can indicate close physical conditions in the extended
components, the radiation of which dominates in the sources with steep spectra.

We determined the Pearson correlation coefficients $r$ for the pairs
``decimal logarithm of the integral flux density--spectral index''.
We considered the combinations between $\alpha_{0.5}$ and the flux densities
at the frequencies of 150\,MHz, 1.4, 3.94 and 4.85\,GHz and, respectively,
for $\alpha_{3.94}$ both for the sources with steep and flat spectra.
The correlation coefficient for the pair $\alpha_{0.5}$ and
$F_{150 MHz}$~(Fig.~\ref{Fig5}a) for the sources with steep spectra amounted to
$r=-0.42$, while for sources with flat spectra is was equal to $r=-0.50$.
Here, as the flux density increases, the radio spectrum gets steeper.
For the other pairs of parameters, the correlation coefficients proved
to be from 0.2 and smaller. Thus, for $\alpha_{0.5}$ and
$F_{1.4 GHz}$ (Fig.~\ref{Fig5}b) $r$ amounts to --0.22, and --0.09 for
the sources with steep and flat spectra, respectively;
for $\alpha_{3.94}$ and $F_{3.94 GHz}$ (Fig.~\ref{Fig5}c) $r=-0.14$,
and $r=-0.15$; for $\alpha_{3.94}$ and $F_{4.85 GHz}$ (Fig.~\ref{Fig5}d)
$r=-0.12$, and $r=-0.13$.
\begin{figure*}[]
\setcaptionmargin{5mm}
\onelinecaptionsfalse
\centerline{
\vbox{
\hbox{
\includegraphics[angle=0,width=0.5\textwidth,clip]{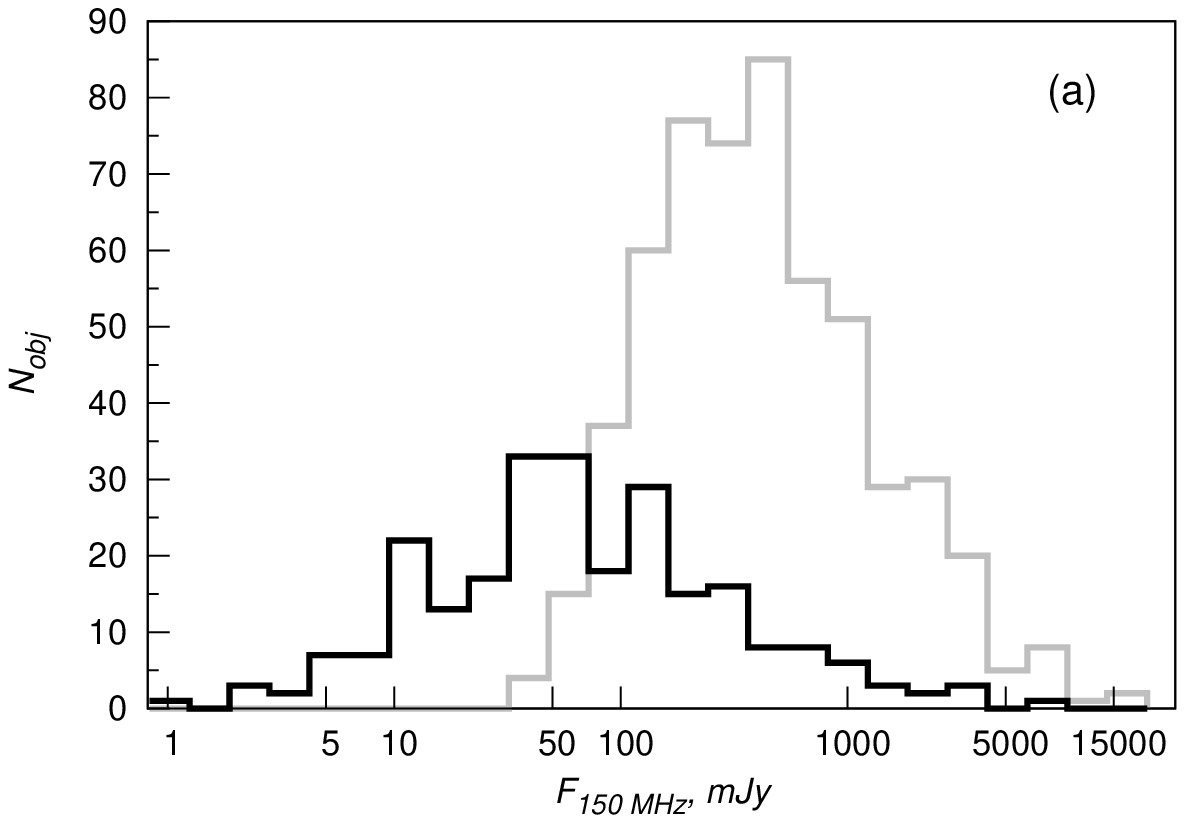}
\includegraphics[angle=0,width=0.5\textwidth,clip]{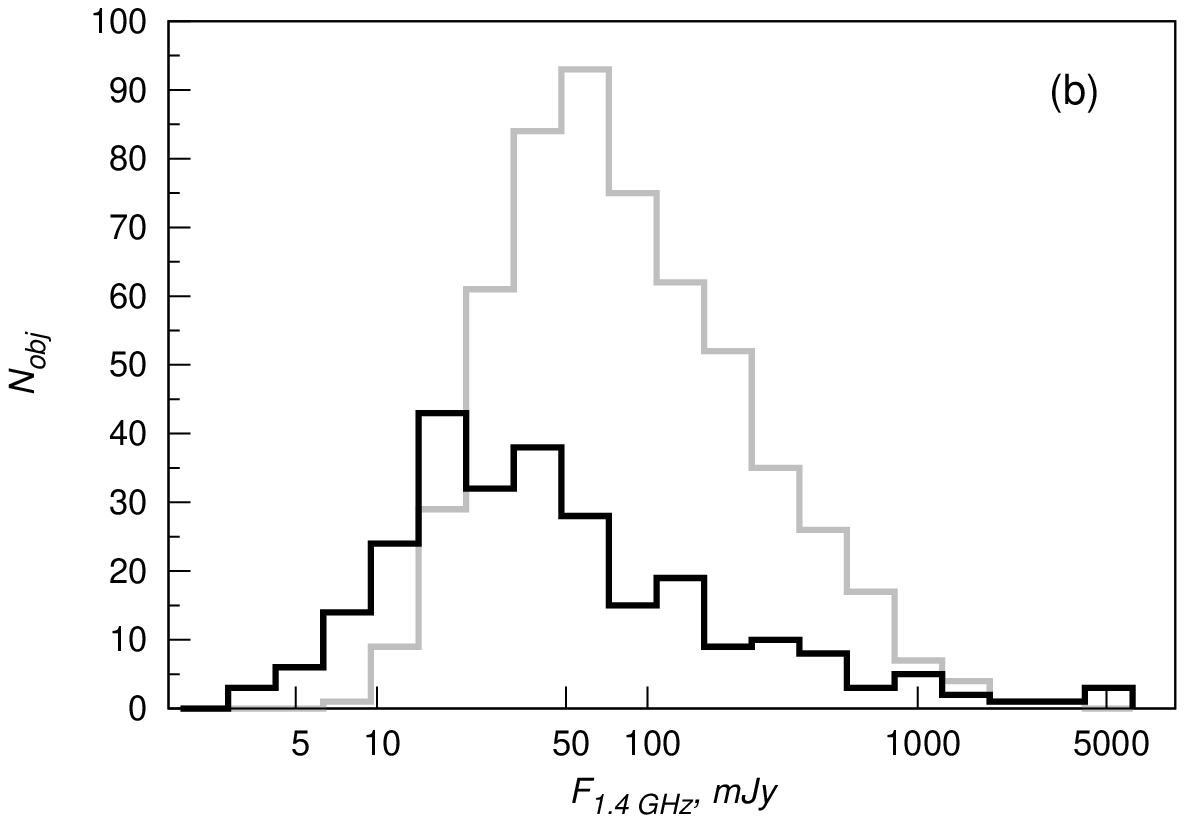}
}
\hbox{
\includegraphics[angle=0,width=0.5\textwidth,clip]{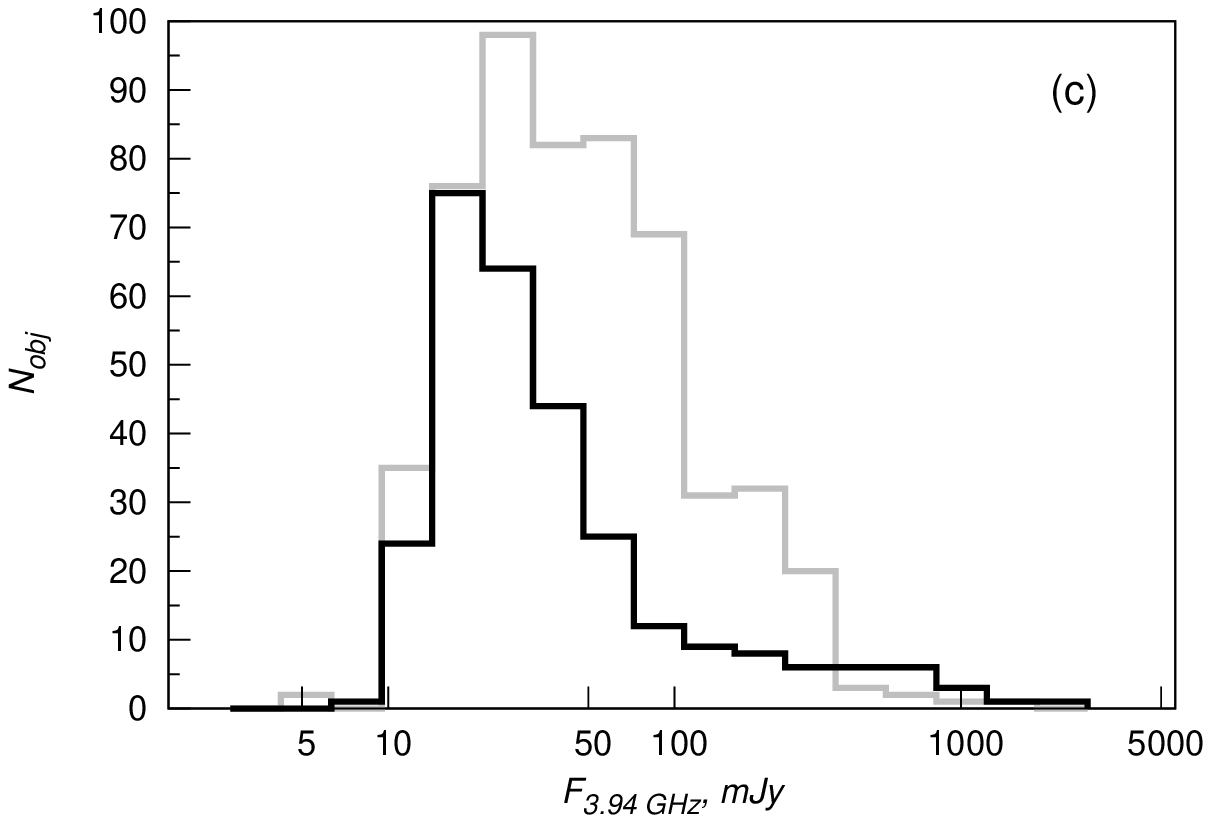}
\includegraphics[angle=0,width=0.5\textwidth,clip]{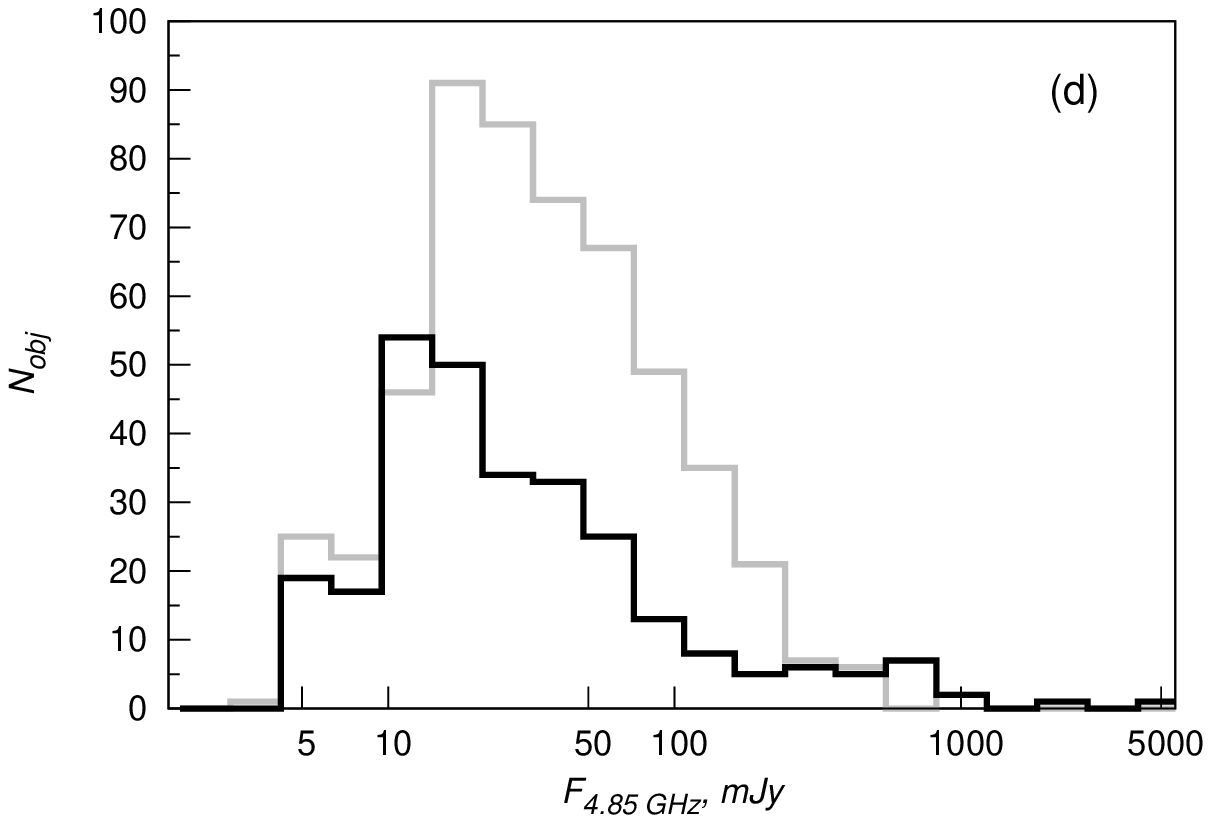}
}
}
}
\caption{
The distribution of the integral flux density for the sources with steep and
flat spectra: grouping by $\alpha_{0.5}$ for 150\,MHz (a) and 1.4\,GHz (b);
grouping by $\alpha_{3.94}$ for 3.94\,GHz (c) and 4.85\,GHz (d).
The gray bars denote the sources with steep spectra, the black line describes
the sources with flat spectra.
}
\label{Fig6}
\end{figure*}

We also compared the distribution of integral flux densities for these two
groups of sources. Figures ~\ref{Fig6}a and \ref{Fig6}b present
the histograms with the distribution of $F_{150 MHz}$ and $F_{1.4 GHz}$, where
the separation into the SS (gray bars) and FS-groups (black line)
is produced by $\alpha_{0.5}$, and Figs.~\ref{Fig6}c and \ref{Fig6}d
represent the histograms with the distribution of $F_{3.94 GHz}$ and $F_{4.85 GHz}$
with a division into the SS and FS-sources by $\alpha_{3.94}$.
In general, the FS-sources are weaker in terms of the flux density than
the SS-sources. It is interesting that in Figs.~\ref{Fig6}c and
\ref{Fig6}d both the maxima of histograms (about 20\,mJy), and
the type of distributions approximately coincide for the SS and FS groups.
At the frequency of 1.4\,GHz, the maximum (Fig.~\ref{Fig6}b) for
the SS-sources is shifted relative to the maximum for the FS-sources and
is located at 40--50\,mJy. A significant difference in the position of
the distribution maximum $F_{150 MHz}$ between the SS and FS-sources is
noticeable at 150\,MHz. We believe that this is due to a different
brightness of the extended components in SS and FS-sources, while
at the frequencies of 3.94--4.85\,GHz, prevailed by the contribution from the
nuclear part, there is no such noticeable difference.

Comparing the diagrams in Fig.~\ref{Fig5} and \ref{Fig6}, we may suggest
that the sources with steep and flat spectra refer to different types
of objects, where the energy physics of the jets is determined by various
types of accretion. To test this assumption, we would have to at least
engage the optical range data for the classification of parent galaxies
of radio sources.

\section{CONCLUSION}

A strip of the sky covered by the surveys of the ``Cold'' experiment
over 1980--1999, by the brightest sources, and the sweep by the right
ascension is $15^h$, which gives the area (with the subtraction
of $2^m$ calibrations) of about $150\square^{\circ}$. Weak sources
($F_{3.94}\le17.5$\,mJy, 30\%) are registered in the $10'$-strip, then
the area of the survey amounts to $40\square^{\circ}$. Accordingly,
for the sources brighter than 17.5\,mJy and weaker than 35\,mJy (30\%)
this is $20'$ and $68\square^{\circ}$, for the sources brighter than 35\,mJy
and weaker than 70\,mJy (20\%) -- $25'$ and $96\square^{\circ}$. The
$20'$ strip captures about three quarters of the total number of objects
of the RCR catalog, or slightly more than 600 sources, which coincides with
the number of sources in~\cite{so3}.

We completed the reduction of a region of the survey strip in
the right-ascension interval \mbox{$2^{h}<R.A.<7^{h}$} and obtained a list
of sources and their characteristics from the averaged scans, where
the sky background was determined being ``smoothed'' by the $80^s$ window.
Now the RCR catalog presents sources from the right-ascension interval
$2^h<R.A.<17^h$, where for each radio source identified by two methods
described above, its position on the scan ($RA$) and the integral
flux density ($F$) at the frequency of 3.94\,GHz are determined.

For each of the objects of the RCR-catalog, radio spectra and spectral
indices at 0.5 and 3.94\,GHz are calculated, and also the two-frequency
spectral indices at 74 and 150\,MHz, 1.4, 3.94 and 4.85\,GHz.
To construct the spectra we have engaged all the known information on
the integral flux densities at different frequencies, available from
the CATS, Vizier and NED databases, as well as the estimated values of flux
densities obtained from the VLSSr, GLEAM, TGSS and GB6 survey maps.
First of all, these estimated values were useful constructing the spectra
of the sources that have data on the flux densities only at two
frequencies: 3.94\,GHz (RCR) and 1.4\,GHz (NVSS). These are mostly
the sources with flux densities of less than 30\,mJy. About 80\%
of them have flat spectra ($\alpha > -0.5$).

According to the results of Soboleva et al.~\cite{so2,so3} in RC sources,
divided by the flux density at 3.94\,GHz into four groups: 5--10\,mJy,
10--20\,mJy, 20--30\,mJy and brighter than 30\,mJy, the median value of the
two-frequency spectral index $\alpha^{1.4}_{3.94}$ amounted to
-0.09, -0.22, -0.45 and -0.65, respectively. For the RCR catalog in
the same source groups, the median values $\alpha^{1.4}_{3.94}$ were
respectively as follows: --0.21, --0.41, --0.51 and --0.61. The tendency of
the spectral index flattening at weaker flux densities for the new
determinations of $F_{3.94}$ over the more extensive observational material
of the 1980--1999 surveys has as well persisted. This may be related with
a decrease in the share of FRII-type sources~\cite{fanaroff}.

We have considered the same six types of integral radio spectra as it was
done by Soboleva et al.~\cite{so1}: $S$, $C^-$, $C^+$, the spectra with
a maximum (MPS, GPS, HFP) and with a minimum (upturn) at the frequencies
of 2--5\,GHz, as well as the spectra of a more complex shape (hill).
It turned out that when adding the GLEAM and TGSS data, the number
of spectra approximated by a straight line decreased from 73\% to 35\%,
the number of $C^-$-spectra increased from 19\% to 29\%, and $C^+$ -- from
2\% to 23\%, the number of sources with spectra, having a maximum -- from
4\% to 10\%, and finally, the number of sources with upturn-spectra increased
approximately from 0.01\% to 3\%.

These variations are related to the observational selection. To account for
it, classifying the sources from our list based on the radio spectrum we are
clearly lacking the data on the flux density in the high-frequency
region from 5\,GHz and above.

To determine the possible relationship between the spectral index and
integral flux density, we calculated the correlation coefficients for
different combinations of spectral indices and flux densities at different
frequencies. A noticeable correlation was detected between $\alpha_{0.5}$
and $F_{150 MHz}$  (Fig.~\ref{Fig5}a) both for the sources with steep spectra,
and for the sources with flat spectra. While the flux density increases,
the radio spectrum gets steeper. However, at other frequencies,
the correlation between the integral flux density and spectral index is absent.

In general, FS-sources are weaker in terms of flux density than
the SS-sources. This difference is especially noticeable at the frequency
of 150\,MHz. We believe that this is linked with a different brightness
of extended components in the SS and FS-sources, while at the frequencies
of 3.94--4.85\,GHz, where the contribution from the nuclear part prevails,
there is no such noticeable difference. We may suggest that the
sources with steep and flat spectra belong to different types of objects with
various jet energies, what can be determined by different types of accretion.
However, the environment and the evolutionary stage of the radio source can
also have an effect. To test this assumption, we would have to attract data
of other ranges for the classification of parent galaxies and the
radio sources themselves.

For a complete analysis of the spectra of objects in the deep decimeter sky
surveys (TGSS, NVSS, FIRST), the sensitivity on the centimeter waves
should be one to two orders of magnitude higher than that in the decimeter
surveys, i.e. not worse than dozens of micro-Jansky. Such a sensitivity in
the centimeter range has so far been realized only on very small areas
of the sky. For this reason, the ``Cold'' experiment survey data obtained
over 1980--2000 still remain relevant.

\begin{acknowledgments}
The study was carried out with the partial support of the RFBR grant
No. 17-07-01367.
The work was accomplished with the support of the Ministry of Education and
Science Of the Russian Federation (state contract 14.518.11.7054).
The research made use of the means of access to the Vizier catalogs,
the SIMBAD database (CDS, Strasbourg, France), as well as
the NASA/IPAC Extragalactic Database (NED), supported by the JPL
of the California Institute of Technology under the contract with NASA.

The authors express their deep gratitude to the referee for the comments
that significantly improved the text of the article.
\end{acknowledgments}

\newpage
\onecolumngrid

\onecolumngrid
\setcaptionwidth{\linewidth}
\setcaptionmargin{0mm} %
\onelinecaptionstrue 
\captionstyle{normal}
\begin{longtable*}{|cllllllllc|}
\hline
\caption{The catalog of radio sources at 7.6 cm in the right ascension interval $2^{h}< R.A.< 7^{h}$\label{Tab1}} \\
\hline
   NVSS             &~$\Delta RA_{1}\pm\sigma$~& $\Delta RA_{2}\pm\sigma$~&~~$F_{1}\pm\sigma$~&~~$F_{2}\pm\sigma$~&~$\alpha_{3.94}$~&~$\alpha_{0.5}$~& Mrph & Flx & Sp.  \\
		    &~~s.ss$\pm$s.ss           &~s.ss$\pm$s.ss            &~~~mJy             &~~~mJy            ~&~               ~&~              ~&      &     &     \\
\hline
\endfirsthead
\caption{(Continue)} \\
\hline
~~(1)              ~&~~~~(2)                   &~~~~(3)                   &~~~~~(4)           &~~~~(5)            &~(6)            ~&~(7)           ~& (8)  & (9) & (10)   \\
\hline
\endhead
\hline
\endfoot
\endlastfoot
020421.77+050130.3 &-0.09$\pm$0.65 &               &~~10.6$\pm$2.0   &                 &~0.65 &-0.44 &    &\#  & +u \\
020638.77+044807.2 &~0.03$\pm$0.76 &~3.10          &~~37.0$\pm$4.0   &~~43.8           &-0.73 &-0.73 &    &    & l  \\
020651.70+044828.6 &               &~2.07$\pm$1.76 &                 &~~31.5$\pm$6.0   &-0.20 &-0.06 &    &E   & -  \\
020704.61+050110.4 &-0.02$\pm$0.25 &~0.16$\pm$0.27 &~~10.2$\pm$2.0   &~~10.0$\pm$2.0   &-1.30 &-0.61 &    &    & -  \\
020801.88+050033.3 &~0.05$\pm$0.30 &~0.20$\pm$0.36 &~~~8.2$\pm$2.0   &~~18.3$\pm$7.5   &-0.15 &~0.08 &    &\#  & -  \\
020912.54+050051.7 &-0.21$\pm$0.20 &~0.24$\pm$0.11 &~~30.8$\pm$2.1   &~~32.3$\pm$2.3   &-1.14 &-0.84 &    &    & -  \\
020921.70+050142.7 &~0.10$\pm$0.35 &~0.13$\pm$0.13 &~~15.1$\pm$3.0   &~~17.4$\pm$3.0   &-0.98 &-0.83 &    &    & -  \\
020931.16+045535.0 &-0.13$\pm$0.31 &~0.98$\pm$0.46 &~~16.6$\pm$2.1   &~~16.1$\pm$2.0   &-0.78 &-0.78 &    &    & l  \\
021336.47+051819.2 &               &~2.92          &                 &~134.2$\pm$3.0   &-0.86 &-0.86 &    &    & l  \\
021449.84+050409.7 &-0.06$\pm$0.35 &~0.42$\pm$0.30 &~~26.7$\pm$2.7   &~~31.6$\pm$3.0   &-0.72 &-0.72 &    &    & l  \\
021906.86+050354.1 &~0.72$\pm$0.25 &~0.71$\pm$0.33 &~~20.2$\pm$2.1   &~~14.5$\pm$1.5   &-0.39 &-0.48 &    &    & +  \\
022019.20+045226.1 &               &~2.13$\pm$1.05 &                 &~~32.3$\pm$3.5   &-0.36 &-0.69 &bR  &F   & h  \\
022032.66+050243.6 &-0.06$\pm$0.28 &~0.29$\pm$0.14 &~~52.5$\pm$3.7   &~~60.7$\pm$4.0   &-0.94 &-0.72 &    &    & -  \\
022046.45+050439.2 &~0.04$\pm$1.00 &~0.59$\pm$0.30 &~~14.7$\pm$4.0   &~~15.6$\pm$5.0   &-1.27 &~0.04 &    &    & -p \\
022141.42+044349.3 &               &~3.12$\pm$0.88 &                 &~151.0$\pm$62.0  &-0.83 &-0.83 &    &B   & l  \\
022218.69+050343.8 &-0.08$\pm$0.44 &-1.22$\pm$0.18 &~~18.2$\pm$3.0   &~~26.8           &-1.14 &-0.90 &bR  &    & -  \\
022220.25+050010.3 &~1.48$\pm$0.44 &~1.04$\pm$0.18 &~~~4.5$\pm$1.5   &                 &-1.06 &-0.70 &bR  &    & -  \\
022416.53+045842.8 &               &-1.81$\pm$0.49 &                 &~~13.6$\pm$3.0   &      &      &mR  &    &    \\
022419.41+045657.3 &~0.00$\pm$0.07 &~0.60$\pm$0.49 &~~15.4$\pm$3.0   &~~19.0$\pm$3.0   & 0.12 &~0.66 &mR  &    & +  \\
022509.74+050837.4 &~1.10$\pm$0.73 &~2.11$\pm$0.05 &~~45.0$\pm$2.7   &~~54.0$\pm$7.0   &-1.17 &-0.90 &    &B   & -  \\
022528.41+045316.2 &-0.10$\pm$0.17 &~0.76$\pm$0.94 &~~20.2$\pm$2.0   &~~33.7$\pm$2.0   &-1.08 &-0.60 &    &    & -  \\
022619.89+044631.5 &~0.29$\pm$0.30 &~0.58$\pm$1.65 &~~76.8$\pm$25.0  &~~54.0$\pm$14.0  &-0.94 &-0.74 &    &    & -  \\
022653.88+045233.4 &               &-0.29$\pm$1.23 &                 &~~24.2$\pm$4.0   &~0.05 &~0.05 &    &    & l  \\
022836.14+045619.2 &~1.50$\pm$0.05 &~0.54$\pm$0.87 &~~14.8$\pm$2.0   &~~17.0$\pm$1.0   &-1.15 &-0.41 &    &    & -p \\
022929.95+045318.0 &~3.29$\pm$0.30 &~1.14$\pm$0.63 &~~19.2$\pm$2.0   &~~29.6$\pm$3.0   &-1.18 &-0.83 &    &    & -  \\
023126.85+045846.4 &~0.06$\pm$0.55 &-0.18$\pm$0.33 &~~12.3$\pm$2.5   &~~13.8$\pm$3.0   &-0.85 &-0.85 &    &    & l  \\
023155.98+050234.4 &-0.10$\pm$0.57 &~0.14$\pm$0.32 &~~17.0$\pm$4.0   &~~18.0$\pm$2.0   &-0.74 &-0.51 &    &v   & -  \\
023331.40+044909.3 &               &~0.33$\pm$1.61 &                 &~~35.7$\pm$7.0   &-0.82 &-0.82 &    &E   & l  \\
023407.16+044642.7 &-0.69$\pm$0.80 &~1.69$\pm$0.34 &~173.3$\pm$17.7  &~271.0$\pm$56.0  &-0.23 &~0.59 &    &s   & -G \\
023546.15+045111.4 &               &-0.36$\pm$0.43 &                 &~~27.5$\pm$3.5   &-0.77 &-0.63 &    &    & -  \\
023840.05+045516.8 &-0.10$\pm$0.50 &~0.23$\pm$0.34 &~~45.5$\pm$6.6   &~~48.9$\pm$5.0   &-0.79 &-0.78 &bR  &    & l  \\
023840.80+045752.3 &~0.65$\pm$0.50 &               &~~12.5$\pm$2.0   &                 &-0.45 &-0.79 &bR  &    & +  \\
023950.49+050042.9 &-0.02$\pm$0.08 &~0.11$\pm$0.11 &~~~9.8$\pm$4.5   &~~19.6$\pm$7.0   &-1.42 &-0.60 &    &s   & -p \\
024309.09+045643.3 &-0.11$\pm$0.50 &~0.00$\pm$0.54 &~~17.6$\pm$4.0   &~~16.3$\pm$3.5   &-0.70 &-0.55 &    &    & -  \\
024322.22+045804.2 &-0.21$\pm$0.57 &-0.02$\pm$0.74 &~~10.0$\pm$2.0   &~~12.7$\pm$1.0   &-0.05 &-0.84 &    &    & +u \\
024430.44+044445.8 &               &~0.72$\pm$1.70 &                 &~~59.2$\pm$1.0   &-0.64 &-0.64 &    &E   & l  \\
024754.12+045414.2 &               &~0.39$\pm$0.51 &                 &~~18.4$\pm$3.5   &~0.31 &-0.44 &    &\#  & +u \\
024816.44+045345.0 &~1.02$\pm$0.70 &~0.53$\pm$0.68 &~~24.4$\pm$12.0  &~~26.5$\pm$12.0  &-0.76 &-0.76 &    &    & l  \\
024939.93+044028.8 &-0.40$\pm$0.40 &~1.48$\pm$3.33 &~139.3$\pm$35.0  &~133.0$\pm$22.0  &-0.96 &-0.68 &    &    & -  \\
025239.26+045840.3 &-0.14$\pm$0.30 &~0.67$\pm$0.16 &~~23.9$\pm$2.6   &~~31.0$\pm$2.6   &-0.67 &-0.44 &    &F   & -  \\
025253.93+050226.0 &-0.07$\pm$0.47 &~0.13$\pm$0.24 &~~16.6$\pm$3.0   &~~21.8$\pm$3.0   &-0.62 &-0.62 &    &    & l  \\
025311.49+050032.2 &~0.68$\pm$0.25 &~0.30$\pm$0.40 &~~~8.7$\pm$1.5   &~~10.3$\pm$2.5   &-0.29 &-0.29 &    &\#  & l  \\
025421.04+045723.9 &-0.30$\pm$0.31 &~0.55$\pm$0.18 &~~15.4$\pm$1.5   &~~17.3$\pm$2.0   &-0.70 &-0.64 &    &E   & -  \\
025630.94+050221.1 &-0.06$\pm$0.53 &~0.22$\pm$0.26 &~~16.0$\pm$2.5   &~~25.5$\pm$2.0   &-1.00 &-0.71 &    &    & -  \\
025647.96+050014.1 &~0.77$\pm$0.32 &~1.31$\pm$0.48 &~~~9.7$\pm$2.0   &~~~8.5$\pm$2.0   &-0.23 &-0.62 &    &    & h  \\
025831.38+045309.0 &               &~1.18$\pm$1.66 &                 &~~29.5$\pm$7.0   &-0.72 &-0.72 &    &    & l  \\
025856.77+050410.4 &~0.96$\pm$0.36 &~0.89$\pm$0.17 &~~14.8$\pm$2.0   &~~18.9$\pm$2.0   &-0.89 &-0.78 &    &    & -  \\
030256.65+045521.1 &~0.21$\pm$0.25 &~0.61$\pm$0.35 &~~45.6$\pm$4.0   &~~58.1           &-1.05 &-0.94 &    &    & -  \\
030321.00+050143.5 &-0.16$\pm$0.42 &~0.21$\pm$0.46 &~~12.8$\pm$3.0   &~~14.8$\pm$1.0   &-0.56 &-0.56 &    &F   & l  \\
030357.72+050240.7 &-0.17$\pm$0.30 &~1.39$\pm$0.66 &~~10.5$\pm$1.5   &~~12.0$\pm$2.0   &~0.04 &-0.71 &    &    & +  \\
030456.91+045640.4 &               &-0.34$\pm$0.81 &                 &~~12.5$\pm$2.0   &~0.29 &~0.29 &    &\#  & l  \\
030546.02+045243.3 &-0.17$\pm$1.12 &~1.92$\pm$1.25 &~~39.5$\pm$3.5   &~~26.4$\pm$12.0  &-0.75 &-0.75 &    &    & l  \\
030626.32+045137.2 &~0.62$\pm$0.38 &~0.84$\pm$0.71 &~~41.9$\pm$16.0  &~~37.8$\pm$6.0   &-0.81 &-0.78 &    &    & -  \\
030656.53+045719.3 &~0.11$\pm$0.20 &~0.53$\pm$0.14 &~~50.3$\pm$5.0   &~~53.8$\pm$8.0   &-0.98 &-0.66 &    &    & -  \\
030726.37+045517.5 &               &-2.94$\pm$1.30 &                 &~~15.8$\pm$2.0   &-0.97 &-0.97 &    &    & l  \\
030733.90+045304.6 &               &~2.42$\pm$1.30 &                 &~~23.5$\pm$2.0   &~0.08 &-0.49 &    &E   & +  \\
030810.14+050226.7 &~2.50$\pm$0.06 &-0.01$\pm$0.32 &~~~9.4$\pm$2.0   &~~12.1$\pm$2.0   &~0.37 &-0.71 &    &F\# & +  \\
030833.98+045409.2 &~1.42$\pm$0.35 &~2.29$\pm$0.75 &~~30.4$\pm$4.0   &~~34.4$\pm$2.0   &-0.83 &-0.58 &    &F   & -  \\
031114.39+050314.6 &-0.75$\pm$1.11 &-0.09$\pm$0.30 &~~26.5$\pm$3.0   &~~26.3$\pm$2.0   &-0.81 &-0.81 &    &    & l  \\
031124.23+050742.7 &~0.72$\pm$0.91 &-0.36          &~~25.6$\pm$3.0   &~~29.2           &-0.71 &-0.57 &    &    & -  \\
031147.96+050802.4 &~0.17$\pm$0.65 &~1.45$\pm$0.31 &~~97.9$\pm$16.0  &~100.1$\pm$16.0  &-1.40 &-1.06 &    &F   & -  \\
031321.84+050452.1 &~0.10$\pm$0.30 &~0.51$\pm$0.41 &~~17.5$\pm$2.0   &~~16.9$\pm$4.0   &-0.97 &-0.97 &    &    & l  \\
031347.01+044724.5 &~1.11$\pm$0.70 &~3.31$\pm$3.40 &~~52.3$\pm$10.0  &~~45.1$\pm$4.0   &~0.91 &~0.91 &    &    & l  \\
031532.21+050721.0 &~1.65$\pm$0.47 &~0.22$\pm$1.00 &~~31.7$\pm$6.5   &~~40.4$\pm$6.5   &-0.78 &-0.78 &d   &    & l  \\
031705.35+045838.2 &-0.27$\pm$0.54 &-0.07$\pm$0.07 &~~11.3$\pm$2.6   &~~13.2$\pm$1.0   &-0.07 &-0.07 &    &\#  & l  \\
031736.52+045545.0 &~2.42$\pm$0.20 &~1.61$\pm$0.78 &~~13.8$\pm$2.0   &~~14.9$\pm$2.0   &-0.15 &-0.25 &    &    & +  \\
031752.52+045452.7 &~0.22$\pm$0.67 &~0.82$\pm$0.20 &~~19.8$\pm$3.0   &~~22.5$\pm$4.0   &~0.65 &-0.51 &    &\#  & +u \\
031841.77+044137.1 &-0.19$\pm$0.36 &-0.50$\pm$0.55 &~200.3$\pm$19.0  &~188.0$\pm$24.0  &-1.07 &-0.93 &db  &B   & -  \\
031844.85+050614.4 &~1.91$\pm$0.76 &~1.47$\pm$0.67 &~~44.2$\pm$4.5   &~~47.9$\pm$7.0   &-0.80 &-0.51 &    &F   & -  \\
031858.07+045914.3 &-0.02$\pm$0.30 &-0.03$\pm$0.11 &~~53.3$\pm$4.0   &~~58.8$\pm$5.0   &-0.33 &-0.92 &    &F   & h  \\
031903.22+045607.9 &~0.08$\pm$0.65 &-0.49$\pm$0.04 &~~24.1$\pm$3.0   &~~30.0$\pm$5.0   &-1.08 &-1.18 &    &\#  & -p \\
031926.47+050448.7 &~0.36$\pm$0.41 &~1.36$\pm$0.42 &~~79.8$\pm$5.0   &~~96.7$\pm$7.5   &-0.73 &-0.66 &    &F   & -  \\
032125.00+045849.6 &-0.19$\pm$0.38 &~0.28$\pm$0.13 &~~15.7$\pm$3.0   &~~14.0$\pm$3.0   &-1.25 &-0.50 &    &F   & -p \\
032314.72+044612.7 &~0.21$\pm$1.30 &~1.02$\pm$0.71 &~117.4$\pm$11.0  &~143.1$\pm$18.0  &-0.07 &-0.81 &    &F   & h  \\
032407.34+044200.2 &-0.41$\pm$0.48 &-0.02$\pm$0.46 &~126.7$\pm$22.0  &~141.5$\pm$25.0  &-1.08 &-0.94 &    &    & -  \\
032456.18+044640.9 &~2.16$\pm$1.08 &~1.65$\pm$1.18 &~~86.2$\pm$8.7   &~~88.1$\pm$12.0  &-1.25 &-0.17 &    &F   & -p \\
032506.09+050110.1 &~0.26$\pm$0.60 &~1.76          &~~11.6$\pm$2.0   &~~14.3           &~1.48 &-0.25 &    &\#  & +u \\
032640.58+051111.6 &-1.18$\pm$0.34 &               &~~23.4$\pm$10.0  &                 &-0.44 &-0.62 &bR  &    & +  \\
032642.23+044650.7 &~0.23$\pm$0.34 &~0.98$\pm$0.98 &~~27.7$\pm$11.0  &~~51.4$\pm$8.5   &-0.78 &-0.78 &bR  &    & l  \\
032724.74+045559.6 &               &~0.82$\pm$0.43 &                 &~~11.5$\pm$2.0   &-0.39 &-0.79 &    &    & +  \\
032825.57+045344.9 &~0.42$\pm$0.43 &               &~~23.1$\pm$1.0   &~~50.3$\pm$6.0   &-0.73 &-0.86 &dbR &    & +  \\
032826.67+045614.3 &~1.52$\pm$0.43 &~1.24$\pm$0.81 &~~23.1$\pm$1.0   &                 &-0.80 &-0.93 &dbR &    & +  \\
032910.98+050336.5 &-0.01$\pm$0.21 &~0.47$\pm$0.22 &~~22.1$\pm$5.0   &~~26.9$\pm$2.0   &~0.02 &-0.81 &bR  &E   & h  \\
032911.02+045817.3 &-0.45$\pm$0.30 &               &~~~8.9$\pm$1.0   &                 &-0.13 &~0.32 &bR  &\#  & -p \\
032917.08+050443.6 &-0.12$\pm$0.25 &-0.04$\pm$0.10 &~~15.1$\pm$2.0   &~~13.6$\pm$3.0   &-0.79 &-0.79 &    &    & l  \\
032935.83+045549.0 &~0.79$\pm$0.23 &~0.25$\pm$1.17 &~~13.8$\pm$2.0   &~~11.6$\pm$3.0   &~0.46 &~0.46 &    &\#  & l  \\
033226.75+045718.7 &~0.02$\pm$0.33 &~0.62$\pm$0.06 &~~29.1$\pm$6.0   &~~23.5$\pm$5.0   &~1.53 &-0.03 &    &v   & +u \\
033510.40+045723.3 &-0.05$\pm$0.15 &~0.45$\pm$0.08 &~~47.3$\pm$3.0   &~~44.7$\pm$2.0   &-0.98 &-0.09 &    &F   & -p \\
033524.20+050038.3 &~0.13$\pm$0.49 &~0.33$\pm$0.11 &~~12.1$\pm$2.0   &~~11.5$\pm$2.0   &~0.17 &~0.17 &    &\#  & l  \\
033613.25+045935.9 &~0.55$\pm$0.64 &~0.57$\pm$0.22 &~~~7.0$\pm$2.0   &~~~8.9$\pm$2.0   &-0.72 &-0.72 &    &B   & l  \\
033726.24+045944.7 &~0.04$\pm$0.23 &~0.65$\pm$0.64 &~~19.8$\pm$2.0   &~~24.0$\pm$5.0   &-0.98 &-0.66 &bR  &    & -  \\
033726.67+045005.5 &               &~0.29$\pm$0.66 &                 &~~87.3$\pm$11.0  &-1.00 &-0.79 &bR  &F   & -  \\
033750.84+045833.2 &~0.84$\pm$0.15 &~0.00$\pm$0.84 &~~~9.8$\pm$1.0   &~~12.1$\pm$1.0   &-0.77 &-0.77 &    &    & l  \\
033901.60+051542.4 &               &~2.49$\pm$0.21 &                 &~~70.0$\pm$7.0   &-0.79 &-0.79 &    &    & l  \\
033959.59+050058.3 &~0.71$\pm$0.12 &~0.77$\pm$0.18 &~~12.1$\pm$1.5   &~~12.8$\pm$2.0   &~0.04 &~0.04 &    &\#  & l  \\
034024.79+045829.8 &~0.01$\pm$0.15 &~0.07$\pm$0.23 &~~21.1$\pm$2.0   &~~22.0$\pm$2.0   &~0.12 &~0.12 &    &s   & l  \\
034041.76+045736.3 &~1.36$\pm$0.28 &-0.52$\pm$0.73 &~~11.2$\pm$2.0   &~~13.5$\pm$5.0   &-0.04 &-0.45 &b   &    & +  \\
034109.80+050709.6 &~1.45$\pm$0.32 &~0.86$\pm$0.61 &~~60.4$\pm$6.0   &~~56.9$\pm$7.0   &-0.97 &-0.86 &    &    & -  \\
034151.93+045925.3 &~0.49$\pm$0.25 &-2.01$\pm$3.03 &~~~8.5$\pm$1.0   &~~~9.0$\pm$1.0   &-0.54 &-0.30 &    &    & -  \\
034231.79+044740.6 &~4.14$\pm$0.58 &~2.16$\pm$1.62 &~~53.7$\pm$6.0   &~~70.8$\pm$2.0   &-0.72 &-0.60 &    &F   & -  \\
034243.15+044527.5 &               &-0.18          &                 &~~82.6           &-0.67 &-0.67 &    &E   & l  \\
034329.99+045750.3 &-0.01$\pm$0.26 &~0.21$\pm$0.23 &1097.5$\pm$50.0  &1133.1           &-0.97 &-0.83 &    &V   & -  \\
034554.43+045729.5 &~0.33$\pm$0.41 &-0.44$\pm$0.70 &~~12.9$\pm$2.0   &~~15.3$\pm$1.0   &-0.39 &-0.60 &    &    & h  \\
034628.75+045545.5 &~0.07$\pm$0.25 &-0.03$\pm$0.76 &~~14.2$\pm$2.0   &~~17.3           &-0.28 &-1.30 &d   &    & h  \\
034656.76+045653.8 &               &-0.10$\pm$0.01 &                 &~~12.3$\pm$3.0   &-0.21 &-0.36 &    &    & +  \\
034824.81+045421.7 &~0.10$\pm$0.61 &~0.82$\pm$0.54 &~~25.6$\pm$5.0   &~~24.3$\pm$3.0   &-0.46 &-0.46 &    &F   & l  \\
034828.10+050151.6 &               &~1.03          &                 &~~12.3$\pm$2.0   &-0.52 &-1.05 &    &    & h  \\
034901.48+051038.4 &~0.32$\pm$0.50 &~0.12$\pm$0.69 &~~55.1$\pm$8.0   &~~66.1$\pm$10.0  &-0.76 &-0.76 &    &B   & l  \\
034931.08+050042.4 &-0.28$\pm$0.36 &-0.09$\pm$0.33 &~~22.3$\pm$4.0   &~~25.0$\pm$1.0   &-0.67 &-0.67 &    &    & l  \\
034940.30+045731.2 &-0.10$\pm$0.10 &~0.03$\pm$0.36 &~~12.8$\pm$4.5   &~~15.8$\pm$4.0   &-0.02 &-0.02 &    &F\# & l  \\
035054.23+050620.9 &~1.09$\pm$0.70 &~0.41$\pm$0.28 &~430.9$\pm$30.0  &~399.4$\pm$27.0  &-0.70 &~0.15 &b   &B   & -p \\
035203.68+044612.0 &               &-1.38$\pm$0.74 &                 &~~59.9$\pm$8.0   &-0.84 &-0.31 &    &    & -p \\
035208.14+045128.5 &~0.05$\pm$0.20 &~2.92          &~~40.3$\pm$7.0   &~~35.6$\pm$5.0   &-0.83 &-0.83 &    &    & l  \\
035303.88+050431.1 &~1.52$\pm$0.22 &~2.14$\pm$0.33 &~~28.6$\pm$6.5   &~~32.5$\pm$1.5   &-0.15 &-0.37 &    &    & +  \\
035424.14+044107.3 &-0.86$\pm$1.10 &~0.53$\pm$1.04 &~193.0$\pm$35.0  &~190.4$\pm$8.0   &-0.13 &-0.43 &    &B   & h  \\
035454.40+050250.2 &-0.25$\pm$0.15 &-0.41          &~~13.8$\pm$2.0   &~~21.1           &-0.97 &-0.81 &    &    & -  \\
035515.52+045703.1 &               &~1.15$\pm$0.56 &                 &~~~9.2$\pm$2.0   &~0.70 &-0.70 &    &\#  & l  \\
035602.18+045602.8 &~0.54$\pm$0.15 &~0.88          &~~11.7$\pm$2.0   &~~15.8           &-0.76 &-0.76 &    &    & l  \\
035659.95+045947.7 &               &-0.82$\pm$0.21 &                 &~~~9.1$\pm$1.5   &-0.10 &-0.69 &    &    & +  \\
035815.51+045449.1 &               &~2.31$\pm$0.05 &                 &~~12.5$\pm$4.0   &~0.55 &-0.76 &    &    & +u \\
040311.59+045929.0 &~0.35$\pm$0.05 &~0.37$\pm$0.93 &~~~9.1$\pm$2.0   &~~~8.2$\pm$3.0   &~0.35 &~0.35 &    &\#  & l  \\
040332.04+045817.3 &-0.07$\pm$0.45 &~0.42$\pm$0.22 &~~45.3$\pm$6.0   &~~45.8$\pm$6.0   &-0.75 &~0.23 &    &    & -p \\
040404.37+045839.5 &~0.12$\pm$0.05 &~0.57          &~~11.0$\pm$2.0   &~~10.4$\pm$2.0   &~0.13 &-0.51 &    &s   & +  \\
040424.21+050633.6 &-2.19$\pm$0.47 &-1.29$\pm$0.03 &~~20.6$\pm$3.0   &~~43.6$\pm$2.0   &-0.82 &-0.82 &b   &    & l  \\
040427.26+050207.2 &-0.04$\pm$0.50 &~0.97          &~~30.5$\pm$4.0   &~~39.4$\pm$2.0   &-0.79 &-0.79 &    &    & l  \\
040626.84+044753.2 &~1.56$\pm$1.10 &-1.82$\pm$4.50 &~~54.4$\pm$10.0  &~~61.4$\pm$10.0  &-0.92 &-0.92 &    &    & l  \\
041034.32+045540.3 &               &~1.19$\pm$0.61 &                 &~~12.6$\pm$4.0   &-0.08 &-0.33 &    &    & +  \\
041319.72+045839.7 &-0.33$\pm$0.56 &~0.18$\pm$0.17 &~~23.4$\pm$3.0   &~~25.2$\pm$3.0   &-0.35 &-0.73 &    &E   & h  \\
041330.97+045247.7 &-1.28$\pm$0.12 &-1.21$\pm$0.67 &~~27.2$\pm$2.0   &~~31.7$\pm$3.0   &-0.60 &-0.60 &    &    & l  \\
041510.24+050144.4 &               &~1.32$\pm$0.56 &                 &~~10.9$\pm$4.0   &-0.64 &-0.76 &    &    & +  \\
041752.68+044404.8 &~1.38$\pm$0.45 &-5.30          &~~59.7$\pm$7.0   &~~66.7$\pm$15.0  &-0.17 &-0.91 &b   &    & h  \\
042003.08+045101.9 &-0.62$\pm$0.44 &~0.54$\pm$0.29 &~~36.6$\pm$6.0   &~~41.8$\pm$6.0   &-0.45 &-0.45 &    &    & l  \\
042154.98+050230.5 &~0.57$\pm$0.28 &~0.26$\pm$0.23 &~~19.1$\pm$2.0   &~~17.8$\pm$2.0   &-0.20 &-0.55 &    &B   & h  \\
042333.58+045451.3 &~0.58$\pm$0.37 &~1.21$\pm$0.32 &~~20.4$\pm$5.0   &~~20.1$\pm$5.0   &-0.69 &-0.69 &    &    & l  \\
042545.15+045028.3 &               &~1.30$\pm$2.50 &                 &~~24.9$\pm$3.0   &-0.49 &-0.49 &    &    & l  \\
042619.18+045025.7 &-0.21$\pm$0.79 &~0.40$\pm$0.40 &~432.5$\pm$54.0  &~434.4$\pm$54.0  &-1.12 &-0.75 &    &    & -  \\
042636.60+051818.0 &               &~7.60          &                 &~375.0$\pm$      &-0.23 &~0.18 &    &B   & -p \\
042747.61+045708.9 &-0.18$\pm$0.44 &~0.46$\pm$0.20 &~645.9$\pm$65.0  &~642.7$\pm$73.0  &-0.26 &-0.26 &    &V   & l  \\
043311.04+052115.4 &~3.89$\pm$0.90 &~0.83$\pm$1.65 &1333.2$\pm$120.0 &1878.0$\pm$200.0 &-0.13 &-0.89 &m   &V   & h  \\
043551.33+045612.6 &~0.95$\pm$0.63 &~1.43          &~~13.2$\pm$3.0   &~~10.4$\pm$3.0   &-0.86 &-0.86 &    &    & l  \\
043558.30+045723.9 &~0.30$\pm$0.25 &~1.36          &~~13.2$\pm$2.0   &~~~9.2$\pm$2.0   &-0.28 &~0.09 &    &\#  & p  \\
043611.99+050127.0 &-0.51$\pm$0.55 &~0.31$\pm$0.38 &~~12.3$\pm$2.0   &~~13.3$\pm$2.0   &-0.66 &-0.66 &    &    & l  \\
043629.74+050034.9 &-0.08$\pm$0.83 &~0.18$\pm$0.56 &~~13.1$\pm$3.0   &~~16.2$\pm$3.0   &-0.30 &-0.30 &    &    & l  \\
043722.65+050529.6 &               &-3.38$\pm$1.38 &                 &~~25.0$\pm$3.0   &-0.64 &-0.64 &    &    & l  \\
043732.83+045139.0 &               &~3.77          &                 &~~11.0$\pm$3.0   &-1.39 &-1.10 &    &    & -  \\
043848.16+044936.2 &               &-0.21$\pm$0.74 &                 &~~46.3$\pm$15.0  &-0.57 &-0.73 &    &    & +  \\
044014.54+050002.9 &~0.20$\pm$1.20 &~1.12$\pm$0.17 &~~14.1$\pm$2.0   &~~15.3$\pm$1.0   &~0.01 &-0.83 &    &    & +u \\
044136.20+045403.4 &~1.00$\pm$0.10 &~1.12$\pm$0.05 &~~19.8$\pm$4.0   &~~19.9$\pm$5.0   &-1.23 &~0.74 &    &\#  & -p \\
044148.48+044848.7 &               &-3.14          &                 &~~28.3$\pm$3.0   &-0.51 &-0.51 &d   &    & l  \\
044417.89+050126.8 &~0.08$\pm$0.35 &~0.32$\pm$0.18 &~~58.2$\pm$6.0   &~~64.6$\pm$5.0   &-1.09 &-0.95 &    &    & -  \\
044455.22+045659.7 &-0.06$\pm$0.48 &-0.05$\pm$0.61 &~~20.0$\pm$3.0   &~~22.0$\pm$7.0   &~0.93 &~0.08 &    &\#  & +u \\
044924.30+045844.5 &~0.21$\pm$0.33 &~0.89$\pm$0.23 &~~10.7$\pm$2.0   &~~11.7$\pm$2.0   &-1.08 &-0.73 &    &    & -  \\
044935.43+050102.3 &~0.29$\pm$1.03 &~1.62$\pm$0.15 &~~13.9$\pm$2.0   &~~11.9$\pm$2.0   &-0.59 &-0.59 &    &    & l  \\
045000.72+051254.9 &~0.03$\pm$2.60 &~2.14          &~~36.5$\pm$3.0   &~~33.2$\pm$5.0   &-0.60 &-0.60 &    &    & l  \\
045110.15+045054.8 &-0.57$\pm$0.30 &~0.08$\pm$0.57 &~~41.5$\pm$4.0   &~~38.3$\pm$14.0  &-0.72 &-0.57 &d   &E   & -  \\
045113.48+043751.2 &               &~2.37          &                 &~~30.6           &-0.88 &-0.58 &    &s   & -  \\
045151.26+050134.7 &-0.52$\pm$0.55 &~1.61          &~~~9.3$\pm$2.0   &~~~9.1$\pm$3.0   &-0.74 &-0.74 &    &    & l  \\
045322.45+051052.6 &~1.29$\pm$0.65 &-0.33          &~~64.8$\pm$12.0  &~~57.2$\pm$8.0   &-0.09 &~0.79 &    &    & -g \\
045503.78+045302.0 &~0.61$\pm$0.27 &~0.56$\pm$0.35 &~~30.9$\pm$5.0   &~~35.3$\pm$7.0   &-1.06 &-0.93 &    &    & -  \\
045544.48+045051.9 &~0.47$\pm$0.62 &~1.26$\pm$0.54 &~~37.0$\pm$5.0   &~~32.8$\pm$4.0   &-0.53 &-0.53 &    &    & l  \\
045754.69+045354.3 &~0.49$\pm$0.39 &~0.93$\pm$0.28 &~~87.8$\pm$5.5   &~~75.7$\pm$7.5   &-0.98 &-0.86 &    &    & -  \\
045815.27+050410.4 &-0.01$\pm$0.60 &~0.43$\pm$0.38 &~~73.6$\pm$5.0   &~~71.3$\pm$7.0   &-1.17 &-0.94 &    &    & -  \\
045905.59+045609.8 &-0.24$\pm$0.26 &-0.18$\pm$0.24 &~~85.7$\pm$8.0   &~~99.8$\pm$8.0   &-0.91 &-0.81 &db  &    & -  \\
050011.77+045838.8 &~0.51$\pm$0.53 &               &~~11.2$\pm$2.5   &                 &-0.53 &-0.53 &    &    & l  \\
050026.57+050433.2 &~0.88$\pm$0.70 &               &~~21.5$\pm$2.7   &                 &-1.23 &-0.69 &    &    & -  \\
050043.12+051155.8 &-2.30$\pm$0.20 &               &~~55.1$\pm$7.0   &                 &~0.23 &-0.51 &    &    & hu \\
050523.20+045942.8 &-0.26$\pm$0.25 &~0.51$\pm$0.12 &~872.0$\pm$70.0  &1000.0$\pm$63.0  &-0.06 &-0.55 &    &V   & h  \\
050625.10+050819.7 &~1.01$\pm$1.06 &~2.02$\pm$0.59 &~~75.2$\pm$12.0  &~~66.4$\pm$6.0   &-0.83 &-0.67 &    &v   & -  \\
050649.14+045101.7 &~0.47$\pm$0.25 &~0.40$\pm$1.33 &~~28.9$\pm$4.0   &~~29.6$\pm$9.0   &-0.23 &-1.03 &    &    & h  \\
050709.01+045520.0 &-0.20$\pm$0.32 &~0.33$\pm$0.01 &~~32.3$\pm$3.0   &~~33.8$\pm$3.0   &-0.88 &-0.81 &    &    & -  \\
050825.45+045155.4 &-1.94$\pm$0.25 &~0.82          &~~22.2$\pm$2.0   &~~29.6           &-0.90 &-0.90 &    &    & l  \\
051006.04+045910.0 &-0.28$\pm$0.18 &~1.18$\pm$0.30 &~~11.9$\pm$2.0   &~~11.7$\pm$2.0   &-0.21 &-0.50 &    &    & +  \\
051018.00+045952.7 &-0.07$\pm$0.37 &-0.47$\pm$2.25 &~~11.7$\pm$2.0   &~~12.9$\pm$1.0   &-0.17 &~0.09 &    &\#  & -  \\
051106.30+045854.5 &-0.37$\pm$0.23 &~0.55$\pm$0.34 &~~15.6$\pm$1.6   &~~15.2$\pm$1.0   &-0.76 &-0.76 &    &    & l  \\
051219.39+045610.8 &~0.82$\pm$0.39 &~0.49$\pm$0.83 &~~13.8$\pm$2.0   &~~17.5$\pm$2.0   &~0.10 &-0.44 &    &    & +  \\
051343.45+045854.7 &-0.18$\pm$0.12 &-0.85$\pm$0.09 &~~31.4$\pm$4.0   &~~33.8$\pm$5.0   &-1.51 &-0.02 &bR  &v   & -p \\
051344.36+050347.3 &~0.41$\pm$0.15 &               &~~14.3$\pm$3.0   &~~13.1$\pm$5.0   &-1.10 &-0.49 &bR  &    & -  \\
051359.03+050235.7 &-0.19$\pm$0.24 &-0.11$\pm$0.48 &~~23.9$\pm$2.5   &~~22.0$\pm$2.0   &-1.45 &~1.67 &    &E\# & -g \\
051539.19+045947.5 &-0.05$\pm$0.20 &~0.26$\pm$0.43 &~~~9.3$\pm$1.0   &~~~9.0$\pm$1.0   &-0.35 &-0.35 &    &    & l  \\
051711.68+050032.6 &-0.13$\pm$0.34 &~0.28$\pm$0.13 &~~38.1$\pm$3.0   &~~44.0$\pm$4.0   & 0.45 &-0.50 &    &s   & h  \\
051909.69+050520.3 &-0.02$\pm$0.11 &~0.49$\pm$0.26 &~~22.4$\pm$2.0   &~~31.0$\pm$7.0   &-0.72 &-0.72 &    &    & l  \\
051923.70+045900.4 &-0.83$\pm$0.15 &               &~~~8.0$\pm$1.5   &                 &~0.28 &-0.49 &    &\#  & +  \\
052035.50+045401.7 &~0.36$\pm$0.37 &~0.82$\pm$0.38 &~~28.4$\pm$3.0   &~~33.0$\pm$4.0   &-0.46 &-0.82 &    &B   & h  \\
052055.49+050654.7 &~2.00$\pm$0.31 &~1.77$\pm$0.75 &~~48.5$\pm$6.0   &~~46.0$\pm$4.0   &-0.95 &-0.78 &    &    & -  \\
052117.03+050728.8 &~0.15$\pm$0.20 &-0.36$\pm$0.67 &~~67.0$\pm$7.0   &~~80.0$\pm$12.0  &-0.92 &~0.37 &    &B   & -g \\
052241.76+045304.3 &~1.05$\pm$0.73 &-0.09$\pm$1.10 &~~24.2$\pm$3.0   &~~21.0$\pm$5.0   &-0.26 &-0.84 &    &    & h  \\
052326.80+045918.6 &-0.40$\pm$0.15 &~0.50$\pm$0.25 &~~11.0$\pm$2.0   &~~16.0$\pm$3.0   &-0.82 &~0.23 &    &\#  & -p \\
052331.28+050844.2 &               &~1.15$\pm$0.81 &                 &~~66.0$\pm$24.0  &-1.06 &-0.87 &    &    & -  \\
052333.28+045827.7 &-0.41$\pm$0.74 &~0.84          &~~17.9$\pm$3.0   &~~19.0$\pm$5.0   &-0.84 &-0.60 &    &    & -  \\
052431.59+050736.6 &               &~1.09$\pm$1.00 &                 &~~39.0$\pm$5.0   &-0.81 &-0.81 &    &E   & l  \\
052502.08+045432.7 &~0.08$\pm$0.27 &~0.70$\pm$0.26 &~~81.7$\pm$8.0   &~~85.0$\pm$9.0   &-0.83 &-0.73 &    &    & -  \\
052719.63+050153.9 &~0.21$\pm$0.37 &~1.93$\pm$0.25 &~~35.8$\pm$3.0   &~~41.0$\pm$3.0   &-0.89 &-0.64 &    &    & -  \\
052801.46+045750.1 &-0.34$\pm$0.23 &~0.32$\pm$0.06 &~~42.1$\pm$3.0   &~~46.0$\pm$1.0   &-0.56 &-0.56 &    &    & l  \\
053207.80+050243.6 &~0.60$\pm$0.10 &~0.31$\pm$0.30 &~~18.1$\pm$2.0   &~~18.0$\pm$3.0   &-0.69 &-0.69 &d   &    & l  \\
053435.41+050342.5 &~0.18$\pm$0.38 &~0.72$\pm$0.23 &~228.0$\pm$20.0  &~240.0$\pm$21.0  &-1.09 &-0.87 &    &    & -  \\
053603.93+050600.6 &~0.68$\pm$0.14 &~1.46$\pm$0.95 &~~23.8$\pm$5.0   &~~29.0$\pm$5.0   &-0.94 &-0.94 &    &    & l  \\
053816.21+045239.5 &~1.04$\pm$0.93 &~2.91$\pm$0.45 &~~25.8$\pm$5.0   &~~27.0$\pm$4.0   &-0.17 &-0.58 &    &    & h  \\
053849.53+050411.5 &-1.22$\pm$0.10 &-0.66$\pm$0.48 &~~23.9$\pm$2.5   &~~30.0$\pm$4.0   &-0.81 &-0.81 &dR  &    & l  \\
053851.42+050309.7 &~0.67$\pm$0.15 &               &~~11.1$\pm$2.0   &                 &      &      &dR  &    &    \\
053957.88+045359.5 &~0.22$\pm$0.10 &~0.80$\pm$0.33 &~~20.9$\pm$4.5   &~~33.0$\pm$5.0   &-0.84 &-0.84 &    &    & l  \\
054118.70+050900.2 &~1.07$\pm$1.05 &~1.77$\pm$0.45 &~~87.3$\pm$8.5   &~100.0$\pm$20.0  &-1.02 &-0.86 &    &    & -  \\
054246.21+045419.6 &~0.15$\pm$0.29 &~0.51$\pm$0.18 &~~58.4$\pm$5.0   &~~58.0$\pm$5.0   &-0.55 &-0.19 &    &s   & -p \\
054405.13+045906.4 &~0.65$\pm$0.24 &~1.66$\pm$0.34 &~~11.4$\pm$2.0   &~~14.0$\pm$1.0   &-0.20 &-0.64 &    &    & +  \\
054555.90+045943.6 &-0.06$\pm$0.27 &~0.75          &~~14.8$\pm$2.0   &~~12.0           &-1.25 &-0.56 &    &s   & -p \\
054948.75+045246.4 &-0.13$\pm$0.76 &~1.09$\pm$1.00 &~~20.4$\pm$4.0   &~~32.0$\pm$9.0   &-0.52 &-0.65 &    &    & +  \\
055256.16+044725.3 &~1.62$\pm$1.07 &~2.99$\pm$0.48 &~~82.3$\pm$18.0  &~~94.0$\pm$2.0   &-1.09 &-0.72 &d   &    & -  \\
055313.77+045549.6 &-0.02$\pm$0.67 &~0.44$\pm$0.31 &~~35.7$\pm$4.0   &~~42.0$\pm$5.0   &-1.16 &-0.13 &    &s   & -p \\
055652.59+050937.2 &~0.55$\pm$1.50 &~1.15$\pm$1.50 &~~49.1$\pm$8.0   &~~46.0$\pm$6.0   &-1.20 &-0.54 &    &    & -p \\
055902.37+045304.8 &~0.12$\pm$2.14 &~0.83$\pm$1.40 &~~20.7$\pm$2.0   &~~28.0$\pm$2.0   &-0.77 &-0.66 &    &    & -  \\
055936.84+045800.8 &-0.53$\pm$0.15 &~0.26$\pm$0.02 &~~14.7$\pm$2.0   &~~16.0$\pm$2.0   &-1.17 &-0.95 &    &    & -  \\
060033.87+045601.1 &~0.04$\pm$1.24 &               &~~11.6$\pm$2.0   &                 &~0.08 &-0.92 &    &    & +  \\
060404.70+045657.4 &~0.28$\pm$0.10 &-0.92          &~~~9.6$\pm$2.5   &~~12.0$\pm$2.5   &-0.87 &-0.32 &d   &\#  & -p \\
060428.72+045958.8 &~0.14$\pm$0.40 &~0.58$\pm$0.77 &~~~8.4$\pm$2.0   &~~12.0$\pm$3.0   &-0.87 &-0.61 &    &    & -  \\
060537.91+050020.5 &               &-1.49$\pm$0.35 &                 &~~11.0           &~0.42 &-0.67 &    &    & +u \\
060612.31+045743.1 &~0.00$\pm$0.27 &~0.21$\pm$0.40 &~~21.2$\pm$4.0   &~~22.0$\pm$3.0   &-0.53 &-0.53 &    &v   & l  \\
060659.73+050659.2 &~1.64$\pm$0.34 &~1.43$\pm$0.40 &~~53.7$\pm$11.0  &~~46.0$\pm$9.0   &-0.81 &-0.71 &d   &v   & -  \\
060715.71+045818.9 &~0.08$\pm$0.62 &-0.53$\pm$0.33 &~~10.3$\pm$2.0   &~~~9.0$\pm$3.0   &-0.74 &-0.74 &    &    & l  \\
060829.14+050115.3 &~0.54$\pm$0.22 &~0.98$\pm$0.16 &~~13.6$\pm$3.0   &~~15.0$\pm$4.5   &-0.85 &-0.57 &d   &    & -  \\
060947.02+045927.9 &-0.09$\pm$0.14 &~0.02$\pm$0.22 &~~10.9$\pm$2.0   &~~11.0$\pm$2.0   &-1.26 &-0.92 &    &    & -  \\
061003.66+045354.1 &~0.40$\pm$0.40 &~0.55          &~~18.1$\pm$4.0   &~~14.0$\pm$3.0   &-0.84 &-0.87 &    &v   & -  \\
061028.84+050025.8 &~0.74$\pm$0.47 &~0.33$\pm$0.43 &~~11.4$\pm$2.0   &~~16.0$\pm$2.0   &-0.91 &-0.75 &    &    & -  \\
061048.06+050504.4 &               &~0.18          &                 &~~15.0$\pm$3.0   &-0.23 &-0.77 &    &    & +  \\
061217.47+045636.7 &               &~0.27$\pm$0.10 &                 &~~16.0$\pm$7.0   &~1.03 & 0.15 &    &\#  & +u \\
061553.63+045650.9 &               &~0.65$\pm$0.90 &                 &~~13.0$\pm$4.0   &-1.26 &-0.79 &    &    & -  \\
061627.92+045312.0 &               &~1.44$\pm$0.50 &                 &~~15.0$\pm$3.0   &-0.06 &-0.48 &    &    & +  \\
061756.20+045824.9 &               &-0.41          &                 &~~~7.0$\pm$3.0   &~0.60 &-0.18 &    &\#  & +u \\
061823.59+050700.1 &               &~1.72          &                 &~~33.0$\pm$7.0   &-0.07 &-0.48 &    &    & +  \\
061900.21+050630.8 &~0.55$\pm$0.60 &~0.91$\pm$0.13 &~321.0$\pm$39.0  &~299.0$\pm$27.0  &-0.81 &~0.04 &b   &v   & -p \\
061909.63+045400.1 &~0.25$\pm$0.32 &~0.76$\pm$0.37 &~~32.5$\pm$5.0   &~~47.0$\pm$1.0   &~0.52 &-0.66 &b   &    & +u \\
061943.49+045748.3 &~0.67$\pm$0.55 &~0.63          &~~11.3$\pm$3.0   &~~12.0$\pm$5.0   &-0.80 &-0.80 &    &    & l  \\
062128.52+045852.2 &-0.21$\pm$0.10 &               &~~23.1$\pm$7.0   &~~37.0$\pm$5.0   &~0.20 &-0.70 &bR  &v\# & -p \\
062130.07+045258.2 &~1.34$\pm$0.10 &               &~~41.6$\pm$5.0   &                 &-0.72 &-0.58 &bR  &s   & -  \\
062152.90+043834.4 &-4.65$\pm$0.28 &~1.06$\pm$1.00 &~361.6$\pm$20.0  &~392.0$\pm$33.0  &-1.01 &-0.80 &    &    & -  \\
062157.68+045606.8 &~0.13$\pm$0.28 &~0.31$\pm$0.14 &~~45.0$\pm$5.0   &~~61.0$\pm$11.0  &-0.79 &-0.55 &    &s   & -  \\
062207.41+045651.1 &-0.06$\pm$0.24 &-0.07$\pm$0.15 &~~35.4$\pm$5.0   &~~25.0$\pm$4.0   &-0.90 &-0.65 &    &v   & -  \\
062310.75+050410.0 &-1.67$\pm$0.27 &-1.86$\pm$0.14 &~~64.4$\pm$21.0  &~~65.0$\pm$11.0  &-0.74 &-0.74 &m   &v   & l  \\
062325.66+045624.1 &~0.02$\pm$0.11 &~0.34$\pm$0.25 &~~11.3$\pm$2.0   &~~12.0$\pm$2.0   &-0.37 &-0.06 &    &\#  & -  \\
062418.85+045701.8 &~0.02$\pm$0.37 &~0.48$\pm$0.23 &~159.0$\pm$10.0  &~165.0$\pm$30.0  &-0.99 &-0.73 &    &    & -  \\
062450.96+050350.0 &~1.24$\pm$0.50 &~0.35$\pm$0.38 &~~22.0$\pm$3.0   &~~22.0$\pm$4.0   &-1.05 &-0.69 &    &    & -  \\
062549.27+045648.2 &               &~0.81$\pm$0.30 &                 &~~28.0$\pm$9.0   &-0.28 &-0.54 &b   &    & +  \\
062741.83+045803.9 &~0.18$\pm$0.37 &~0.28          &~~25.1$\pm$3.0   &~~52.0$\pm$      &-0.98 &-0.82 &    &v   & -  \\
063605.69+043240.5 &~1.59$\pm$0.15 &               &~477.3$\pm$40.0  &                 &-0.79 &-0.79 &    &    & l  \\
063759.26+045505.5 &~0.71$\pm$0.73 &~0.22$\pm$0.67 &~~15.1$\pm$2.0   &~~18.0$\pm$2.0   &~0.77 &~0.40 &    &\#  & +  \\
063826.31+045246.6 &-0.03$\pm$0.70 &-0.13$\pm$0.94 &~~19.7$\pm$3.5   &~~27.0$\pm$4.5   &-0.79 &-0.79 &    &    & l  \\
063929.62+045937.0 &~0.51$\pm$0.38 &-0.02$\pm$0.27 &~~12.4$\pm$3.0   &~~17.0$\pm$4.0   &-0.26 &-0.57 &    &v   & +  \\
064054.67+050550.3 &-0.62$\pm$0.34 &~0.67$\pm$0.01 &~~38.9$\pm$5.0   &~~36.0$\pm$4.0   &-0.44 &-0.78 &    &    & +  \\
064116.31+044748.5 &~0.28$\pm$0.51 &~0.99$\pm$0.71 &~~61.5$\pm$6.0   &~~59.0$\pm$18.0  &-1.22 &-0.66 &    &    & -  \\
064415.38+050641.5 &~0.33$\pm$0.71 &~1.23$\pm$0.40 &~112.0$\pm$20.0  &~119.0$\pm$5.0   &-0.93 &-0.83 &    &    & -  \\
064753.44+050456.5 &               &-0.22$\pm$0.13 &                 &~~24.0$\pm$3.0   &-0.15 &~1.18 &    &E\# & +u \\
065110.86+045356.1 &~0.81$\pm$0.73 &~0.00          &~~12.2$\pm$2.0   &~~17.0$\pm$6.0   &~0.21 &-0.63 &    &    & +  \\
065327.45+050319.2 &~0.10$\pm$0.53 &               &~~45.5$\pm$6.0   &                 &-0.90 &~0.03 &d   &    & -p \\
065327.47+050851.6 &~0.12$\pm$0.53 &~0.24$\pm$0.46 &~151.0$\pm$15.0  &~149.0$\pm$16.0  &-0.48 &-0.35 &bR  &B   & -  \\
065529.90+045510.9 &~0.13$\pm$0.38 &~0.29$\pm$0.23 &~~42.8$\pm$4.0   &~~48.0$\pm$6.0   &-0.03 &-0.03 &    &s   & l  \\
065848.74+045522.0 &~0.71$\pm$0.55 &               &~~25.1$\pm$1.0   &                 &~0.25 &-1.00 &bR  &v1  & +u \\
065850.15+050206.7 &-0.11$\pm$0.73 &~1.10$\pm$0.71 &~~41.0$\pm$18.0  &~~40.0$\pm$7.0   &~0.00 &~0.00 &bR  &v   & l  \\
065929.43+045603.8 &~0.29$\pm$0.30 &-0.94$\pm$0.92 &~~13.5$\pm$4.0   &~~13.0$\pm$2.0   &~0.28 &~0.28 &    &v\# & l  \\
\end{longtable*}

\end{document}

%% file: sao_cmd_author.tex
\def\saoname{Special Astrophysical Observatory,  Russian Academy of Sciences,
              Nizhnii Arkhyz, 369167 Russia}

\def\squareforqed{\hbox{\rlap{$\sqcap$}$\sqcup$}}

\def\sq{\ifmmode\squareforqed\else{\unskip\nobreak\hfil
\penalty50\hskip1em\null\nobreak\hfil\squareforqed
\parfillskip=0pt\finalhyphendemerits=0\endgraf}\fi}

\def\arcsec{\hbox{$^{\prime\prime}$}}

\def\utw{\smash{\rlap{\lower5pt\hbox{$\sim$}}}}

\def\udtw{\smash{\rlap{\lower6pt\hbox{$\approx$}}}}

\def\diameter{{\ifmmode\mathchoice
{\ooalign{\hfil\hbox{$\displaystyle/$}\hfil\crcr
{\hbox{$\displaystyle\mathchar"20D$}}}}
{\ooalign{\hfil\hbox{$\textstyle/$}\hfil\crcr
{\hbox{$\textstyle\mathchar"20D$}}}}
{\ooalign{\hfil\hbox{$\scriptstyle/$}\hfil\crcr
{\hbox{$\scriptstyle\mathchar"20D$}}}}
{\ooalign{\hfil\hbox{$\scriptscriptstyle/$}\hfil\crcr
{\hbox{$\scriptscriptstyle\mathchar"20D$}}}}
\else{\ooalign{\hfil/\hfil\crcr\mathhexbox20D}}%
\fi}}

\newcommand{\ab}{Astrophysical Bulletin }

\newcommand{\aaa}{Astron. and Astrophys. }

\newcommand{\aas}{Astron. and Astrophys. Suppl. }

\newcommand{\aar}{Astron. Astrophys. Rev. }
\newcommand{\aj}{Astron.~J. }
\newcommand{\apjs}{Astrophys.~J. Suppl. }

\newcommand{\mnras}{Monthly Notices Royal Astron. Soc. }

